\documentclass[
prc,
aps,
superscriptaddress,
twocolumn,
secnumarabic,
balancelastpage,
longbibliography,
amsmath,
amssymb,
nofootinbib,
floatfix,
reprint
]{revtex4-2}

\usepackage{graphicx}      
\usepackage{bm}            
\usepackage{amsmath}
\usepackage{amssymb}
\usepackage[version=3]{mhchem}
\usepackage[colorlinks=true]{hyperref} 
\usepackage{textcomp}

\usepackage{tabularx}

\usepackage{subfig}
\usepackage{float}

\newcommand{\subfigimg}[4][,]{%
  \setbox1=\hbox{\includegraphics[#1]{#3}}
  \leavevmode\rlap{\usebox1}
  \rlap{\hspace*{#4pt}\raisebox{\dimexpr\ht1-2\baselineskip}{#2}}
  \phantom{\usebox1}
}

\let\subfloat\relax

\let\captioncaption\caption
\let\caption\captioncaption

\newcommand\Tstrut{\rule{0pt}{2.6ex}}         
\newcommand\Bstrut{\rule[-0.9ex]{0pt}{0pt}}   

\begin{document}
\title{Secondary Neutron Production from Thick Target Deuteron Breakup}
\author{Jonathan T. Morrell}
\affiliation{Department of Nuclear Engineering, University of California, Berkeley, Berkeley, CA 94720, USA}
\affiliation{Los Alamos National Laboratory, Los Alamos, NM 87545, USA}

\author{Andrew S. Voyles}
\affiliation{Department of Nuclear Engineering, University of California, Berkeley, Berkeley, CA 94720, USA}

\author{Jon C. Batchelder}
\affiliation{Department of Nuclear Engineering, University of California, Berkeley, Berkeley, CA 94720, USA}

\author{Joshua A. Brown}
\affiliation{Department of Nuclear Engineering, University of California, Berkeley, Berkeley, CA 94720, USA}

\author{Lee A. Bernstein}
\affiliation{Department of Nuclear Engineering, University of California, Berkeley, Berkeley, CA 94720, USA}
\affiliation{Nuclear Science Division, Lawrence Berkeley National Laboratory, Berkeley, CA 94720, USA}

\date{\today}
\setlength{\parskip}{1em}

\begin{abstract}
Thick target deuteron breakup is a variable-energy accelerator-based source of high-energy neutrons, with applications in fundamental and applied nuclear science and engineering.  However, the breakup mechanism remains poorly understood, and data on neutron yields from thick target breakup remains relatively scarce.  In this work, the double-differential neutron yields from deuteron breakup have been measured on a thick beryllium target at $\epsilon_d=33$ and 40 MeV, using both time-of-flight and activation techniques.   We have also introduced a simple hybrid model for the double-differential deuteron breakup cross section, applicable in the $\epsilon_d=10$--$100$ MeV energy range on light ($Z\leq 6$) targets.  This model features four empirical parameters that have been fit to reproduce experimental breakup measurements on beryllium targets, using the method of least-squares.  It was shown that these parameters extrapolate well to higher energies, and to other low-Z target materials.  We also include optimization of the parameters that modify the Kalbach systematics for compound and pre-equilibrium reactions, in order to better reproduce the experimental data for beryllium targets at large angles.
\end{abstract}

\maketitle

\section{Introduction}

Thick target deuteron breakup is one of the most intense accelerator-based sources of neutrons in the range of $\approx$10--100 MeV, with the potential for numerous scientific and industrial applications.  At a deuteron energy of $\epsilon_d=40$ MeV on a beryllium target, approximately 9\% of the incident deuteron beam breaks up, and is ``converted" into a forward-focused neutron beam.  Another appealing characteristic of deuteron breakup is that the average energy of neutrons emerging from the reaction is approximately half of the incident deuteron energy.  The fact that the energy and intensity of the outgoing neutron distribution are (approximately) proportional to the energy and intensity of the incident deuteron beam means that the neutron spectrum can be tuned to suit the requirements of a particular application.

However there is a pressing need for improving the characterization of the neutron energy and angle distributions from thick target deuteron breakup.  Many of the available literature data are not of sufficient quality for the applications desired, and are often inconsistent with one another.  In addition, most of the modeling efforts associated with deuteron breakup are either focused on (d,p) transfer reactions in thin targets, rather than on thick target neutron yields, or are simply inaccurate over the energy range or target materials which are useful for applications.

In this work, we present a parameterized, hybrid model for deuteron breakup that has been fit to a selection of literature data on thick beryllium targets, spanning an energy range of $\epsilon_d = 16$--50 MeV.   The model was shown to extrapolate well to higher energies, and to other targets in a similar mass range.  We also present new measurements of the double-differential neutron spectra acquired at $\epsilon_d = 33$ and 40 MeV, which were performed using the time-of-flight and foil activation techniques.

\subsection{Applications of Deuteron Breakup}

The tunable energy spectrum and relatively high intensity of thick-target deuteron breakup make it an attractive neutron source for a variety of applications.  One application that we will examine in this work is isotope production, where breakup has the potential to play a unique role in comparison to other neutron sources.  Nuclear reactors are very intense neutron sources, but generally have a fixed energy spectrum that is well-suited for (n,$\gamma$) isotope production.  Neutron generators (both deuterium-deuterium and deuterium-tritium) are a relatively low-cost source of high energy (e.g. 2.5, 14 MeV) neutrons, but the energy is fixed and the neutron intensity is comparably low.  Spallation neutron sources have a very high intensity, however they emit a very broad range of neutron energies, only a small selection of which are likely useful for isotope production \cite{mosby2016determination}.  

In comparison, the spectrum from a deuteron breakup neutron source could be optimized for a specific (n,p), (n,2n), (n,3n) or (n,$\alpha$) reaction, selectively populating one isotope with a relatively high radiopurity.  Also, because there is a strong energy-angle correlation in the breakup spectrum, different isotope production targets could be arranged at different angles, enabling simultaneous radionuclide production.  In this work, we will present cross section measurements for the production of the \ce{^{64}Cu}, \ce{^{67}Cu}, \ce{^{44}Sc}, and \ce{^{47}Sc} medical isotopes using a deuteron breakup source; however, many other isotope production pathways are possible.

There are also many potential scientific applications of deuteron breakup.  Because the neutron production is coupled to the incident deuteron beam, a pulsed accelerator (such as a cyclotron or linear accelerator) can be used to perform nuclear physics measurements using the time-of-flight (ToF) technique.  This is useful for measuring reaction cross sections as a function of the incident neutron energy, for which there is a paucity of data in the fast neutron range of several MeV \cite{bernstein2019our}.  If the deuteron beam is scaled up significantly in intensity, the spectrum is suitable for neutron damage studies of fusion reactor materials \cite{EHRLICH199872}, or for electronics damage and human dose considerations for space exploration \cite{norbury2014space}.  While not a direct application of deuteron breakup, modeling the breakup reaction is important for interpreting (d,p) reaction data, often used in surrogate nuclear reactions for targets that are either difficult to acquire, or have a low reaction cross section \cite{potel2017toward}.  There are also potential applications for active interrogation studies, or neutron induced transmutation of nuclear waste \cite{ducasse2016investigation}.

\subsection{Background}

Deuteron breakup has a long history of study, beginning in the 1930's when Oppenheimer proposed that deuterons having a kinetic energy larger than their nuclear binding energy ($E_d = 2.225$ MeV \cite{Wang_2021}) could be disintegrated through Coulombic interactions in matter \cite{oppenheimer1935disintegration}. In the 1940's, measurements by Helmholtz, McMillan and Sewell \cite{helmholz1947angular} of uranium target bombardment using high energy deuterons showed that neutrons from this process were emitted at approximately half the incident deuteron energy, and that this process was very forward-focused.  Dancoff and Serber proposed two competing theories explaining these observations, the former based on Coulomb excitation of the deuteron \cite{dancoff1947disintegration}, and the latter being a nuclear process in which the proton is ``stripped" away from the neutron \cite{serber1947production}.  Dancoff also showed that for low-Z nuclei, the Coulomb breakup cross section would be much smaller than for the stripping process.  This assertion is also maintained by more recent theoretical work \cite{nakayama2016theoretical}, and is fairly intuitive as the cross section for Coulomb excitation is proportional to $Z^2$.

The theories of Dancoff and Serber represent the two competing reaction mechanisms responsible for deuteron breakup.  The Dancoff theory of Coulomb excitation is one form of \textit{elastic breakup} (EB), where no excitation energy or particles are transferred to the target nucleus. In this case the breakup is not induced by a nuclear reaction, however nuclear-elastic breakup is also possible.  The Serber theory of proton stripping is an example of a \textit{nonelastic breakup} (NEB) reaction, in which either excitation energy, a particle, or both are transferred to the target nucleus, which necessarily occurs through a nuclear interaction.

The essence of the Serber stripping theory is, as summarized by Potel \textit{et al.} \cite{potel2017toward},  ``the product of the square of the Fourier transform of the ground state wave function of the [deuteron] projectile and the total reaction cross section of the unobserved fragment", which is the proton in this case.  More simply, it assumes that the neutron momentum distribution before and after the interaction, in which the proton is stripped away from the deuteron, remains the same.  This reproduces the basic kinematic behavior experimentally observed for breakup on light targets: that the neutron emerges with an average energy of one-half that of the incident beam, and the angle distribution becomes more forward-focused as the incident energy is increased.

A more rigorous treatment of this process was given using direct reaction theory, in the works of Ichimura, Austern, and Vincent (IAV) \cite{ichimura1985equivalence} using the post-form of the  distorted-wave Born approximation (DWBA) and Udagawa and Tamura (UT) \cite{udagawa1981derivation} using the prior DWBA form.  These theories preserve the essentials of the Serber model, but account for diffraction effects caused by interactions of the elastic breakup channel with the target nucleus \cite{potel2017toward}.  More modern approaches make use of three-body descriptions of breakup, or will perform coupled-channel calculations which are capable of computing multi-step reaction processes to high orders of accuracy \cite{FRESCO}.  All of these described methods make use of optical-model potentials for the deuteron-target interacting system, the parameters of which can be tuned to reproduce experimental results.  Other recent efforts in breakup modeling include semi-empirical formulae fitted to experimental data, such as the works of Kalbach \cite{kalbach2017phenomenological} and Avrigeanu \cite{avrigeanu2017additive}.  However these models have been shown to have poor agreement with neutron yield data on light targets \cite{nakayama2021jendl}, due to the fact that they were mostly fit to (d,p) reaction data on medium-mass nuclides.  This served as the motivation for development of the hybrid breakup model presented in this work, which is specifically tuned to reproduce neutron yield data on low-Z targets, with an eye towards using the forward-focused neutrons as a beam in their own right.

In addition to EB and NEB, there are other nonelastic reaction processes which may contribute to neutron production.  These are compound nucleus formation, pre-equilibrium, and direct reactions leading to excited states in the residual nucleus.  It is argued that due to the different time scales for these reaction mechanisms, they can be treated as an incoherent summation, such that the resulting spectra from each mechanism can simply be added together \cite{nakayama2016theoretical}.  On this basis, the modeling of these processes are generally performed independently from breakup, and can be calculated with a standard nuclear reaction model code (TALYS, EMPIRE, CoH, etc.) \cite{TALYS, HERMAN20072655, CoH3}.

As far as experimental measurements of breakup are concerned, most effort has been focused on (d,p) reaction data on medium-mass targets, often with the goal of providing nuclear structure insight.  This is motivated by two factors.  Experimentally, protons can be detected with much higher efficiency and energy resolution than neutrons, via the use of silicon charged particle detectors.  This enables measurements on rare or enriched target materials, that may not be available in quantities larger than a few milligrams.  The superior energy resolution means that more details of the underlying nuclear physics may be revealed, as reactions to specific states can be observed.  The other motivating factor is that (d,p) reactions can act as a surrogate to (n,$\gamma$) reactions, which are important for a wide variety of applications \cite{arcones2017white}.  The interpretation of these surrogate measurements is one of the principal motivations behind the development of more comprehensive theories for deutron breakup, such as the work of Potel \textit{et al.} \cite{potel2017toward}.

This work is focused on the  development of a predictive model for non-spallation based neutron production in the range of $\epsilon_d = 10$--200 MeV.  As a result, the most valuable data are for low-Z targets which have the lowest overall $dE/dX$, and therefore the longest effective range.  Within this scope, most of the literature measurements are from the 1970's on thick beryllium targets in the 16--55 MeV deuteron energy range, which were motivated by the desire to use fast neutrons in cancer therapy \cite{meulders1975intensity}.  Unfortunately, many of these measurements are discrepant with one another, and most were only performed at forward angles.  This motivated us to perform new, well-quantified measurements at multiple energies and angles.

\section{Parameterization of the Hybrid Breakup Model}

The majority of reported experimental neutron spectra from deuteron breakup are neutron yields from thick targets.  More specifically, they are the neutron production cross sections integrated over the entire range of the deuteron for a given incident energy, rather than energy differential measurements.  This includes neutron production from inelastic reactions other than breakup.  While having thick target yield data is convenient from the applications perspective, using deuteron breakup as a neutron source, it means that forward modeling is required to extract the neutron production cross sections as a continuous function of deuteron energy ($\epsilon_d$).  The total double-differential neutron yields from a thick target can be calculated according to:

\begin{equation}
\frac{d^2Y(E_n, T_e, \theta)}{dE_n,d\Omega} = \frac{\rho_N}{e} \int_0^{E_d} \tau(\epsilon_d) \frac{d^2\sigma(\epsilon_d, \theta)}{d\Omega dE_n} \big(\frac{d\epsilon_d}{dx}\big)^{-1} d\epsilon_d ,
\label{eq:yields}
\end{equation}

\noindent where $\rho_N = \frac{\rho \cdot N_A}{M}$ is the number density of the target, $\tau(\epsilon_d)$ is a parameter representing the transmission of the deuteron beam from the incident energy $E_d$ down to $\epsilon_d$, $\frac{d\epsilon_d}{dx}$ is the deuteron stopping power in the target, and $\frac{d^2\sigma(\epsilon_d, \theta)}{d\Omega dE_n}$ is the double-differential neutron production cross section that we are attempting to unfold.  At very high energies, we would also need to account for neutron attenuation within a necessarily thicker target, as well as tertiary neutrons produced from the secondary proton flux.  However, these contributions are likely to be quite small at the energies relevant to this work.

In order to build a successful model for the neutron production cross sections $\frac{d^2\sigma(\epsilon_d, \theta)}{d\Omega dE_n}$, we must perform an iterative procedure in which the cross sections are predicted, the thick target yields are calculated from these, and then the model is adjusted to better reproduce the yield data.  To perform this optimization, the model requires adjustable parameters.  While one could adjust optical model parameters in a DWBA calculation, we have instead opted to use a parameterzied version of the Serber theory \cite{serber1947production}.  In addition to the relative simplicity of the Serber theory, there are two motivating factors for using this model.  One is that it has already been shown that for low-Z target nuclei (relevant to neutron production targets), the elastic breakup component is almost negligible \cite{nakayama2016theoretical}, such that there is little improvement in accuracy expected from a DWBA calculation.  The exception to this is at low energies ($<15$ MeV), however at these energies, the short range of the deuteron limits the resulting neutron flux, making it generally outside the scope for an accelerator-driven neutron source.  The other advantage of the Serber model is that it does not require a re-tuning of optical model parameters for each target nucleus \cite{pereslavtsev2008evaluation}, which would likely be required for DWBA calculations as global optical model potentials (OMPs) such as the Koning-Delaroche OMPs \cite{koning2003local} are not valid for light nuclei such as \ce{^{6,7}Li} and \ce{^{9}Be}.

With this in mind, we will first describe the parameterized breakup model, and then discuss the selected parameter adjustments and procedures for fitting to experimental data. In the following section, the model parameters have been re-normalized such that the fit coefficients (to be described presently) $c_1$--$c_6$ are equal to unity, according to a (least-squares) fit to selected literature data. These encompass our ``recommended" parameter values, however, as will be shown in section \ref{measurements_section}, this relatively straightforward parameterization enables fitting to integral yield data (activation) as well as enabling fine-tuning of the model to specific energy regions.

As mentioned previously, due to the different time scales of the various neutron producing mechanisms involved, we can divide the cross section into an incoherent sum of the following inclusive cross sections:

\begin{equation}
\frac{d^2\sigma}{d\Omega dE_n} = \frac{d^2\sigma_{BU}}{d\Omega dE_n} + \frac{d^2\sigma_{CM}}{d\Omega dE_n} + \frac{d^2\sigma_{PE}}{d\Omega dE_n} ,
\end{equation}

\noindent where the subscripts $BU$, $CM$ and $PE$ refer to breakup, compound-fusion evaporation reactions, and pre-equilibrium reactions.  Direct reactions were excluded from this study because they only contribute significantly to the neutron yields at low energy ($E_d<10$ MeV), which is outside the scope of our focus on applications such as isotope production, where the expected (deuteron) energy ranges may be on the order of $E_d=30$--$60$ MeV.

For the breakup cross section, it was assumed that the energy and angle distributions were independent, and therefore the cross section could be described separately by a total breakup cross section $\sigma_{BU}$, and probability distributions for the outgoing neutron energy ($P_{BU}(E_n)$) and angle ($P_{BU}(\theta)$) according to:

\begin{equation}
\frac{d^2\sigma_{BU}}{d\Omega dE_n} = \sigma_{BU}(\epsilon_d)P_{BU}(E_n)P_{BU}(\theta) ,
\end{equation}

According to the Serber model, the stripping cross section, $\sigma_{BU}$, should be proportional to the target nuclide radius, approximated as $R=r_0A^{1/3}$, where $r_0 = 1.25\times 10^{-10}$ m.  However, Serber did not include any dependence on the incident deuteron energy.  Therefore, we have combined the Serber formula with the energy dependence from the semi-empirical breakup model by Kalbach \cite{kalbach1988systematics} to obtain:

\begin{equation}
\sigma_{BU}(\epsilon_d) =  57.2\cdot \frac{(A^{1/3}+2^{1/3})}{1+e^{(22.3-\epsilon_d)/\eta_{BU}}} \text{ (mb)} ,
\end{equation} 

\noindent where $\eta_{BU}=9.4$ MeV is a slope parameter, that will be used in the fitting procedure described in section \ref{literature_fitting}.

\begin{figure}[htb]
\centering
\includegraphics[width=0.99\linewidth]{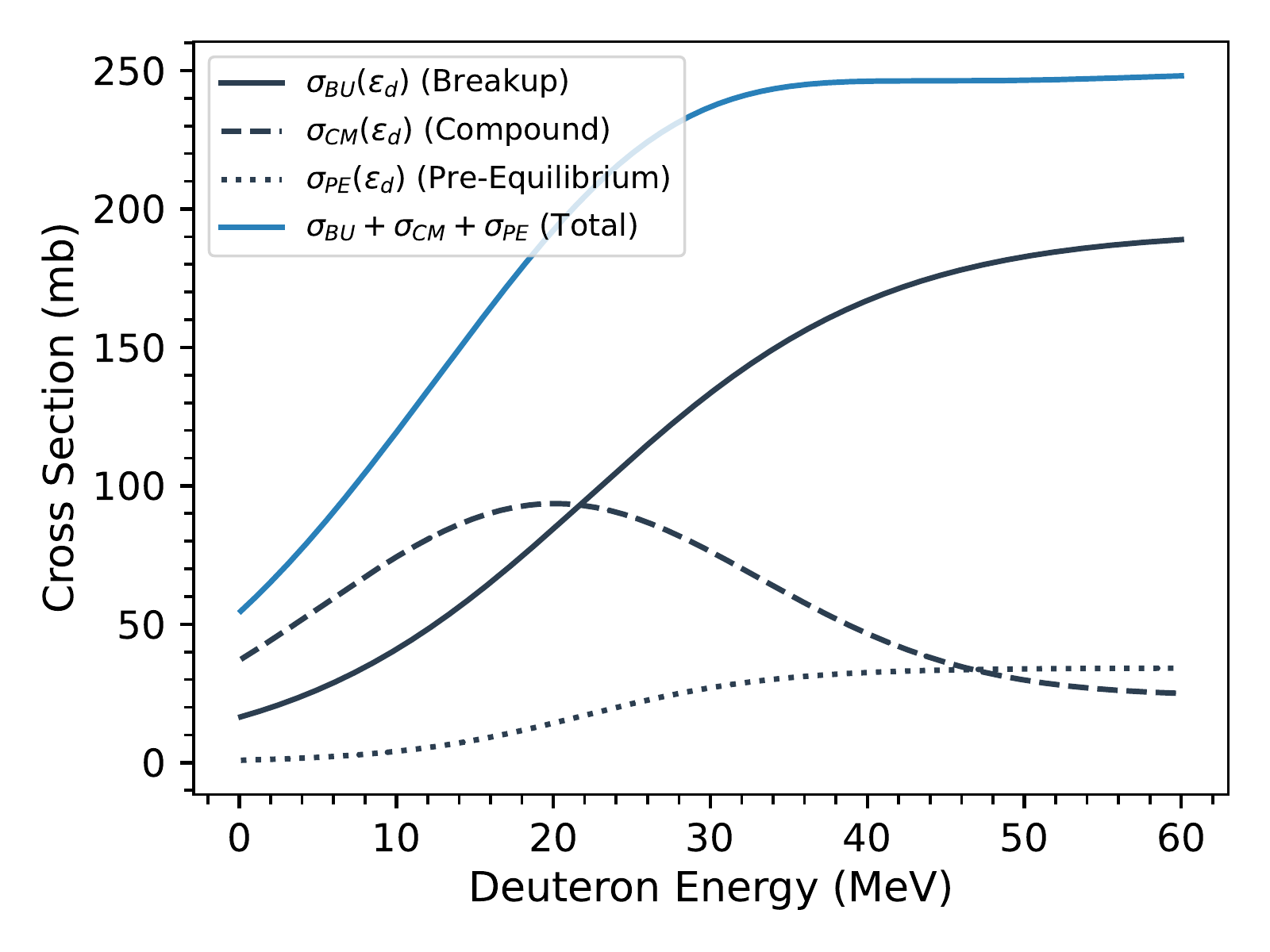}
\caption{The total neutron producing cross sections for the deuteron breakup, compound (evaporation) and pre-equilibrium reaction components.
}
\label{fig:background_xs}
\end{figure}

A comparison of the total breakup cross section with the neutron production cross sections from pre-equilibrium and compound reactions can be seen in Fig. \ref{fig:background_xs}.  These results show that above approximately 20 MeV, breakup is the dominant contributor to the total (angle-integrated) neutron production.

The neutron energy and angle distributions for breakup were taken from the opaque nucleus approximation in the Serber theory.  However the characteristic  widths of these distributions were modified to fit existing literature data.  The resulting energy probability distribution is given by:

\begin{equation}
P_{BU}(E_n) = \frac{N_E\cdot (\epsilon_d - E_c)\cdot w_d}{\pi\Big[\big(E_n - \frac{1}{2}(\epsilon_d - E_c)\big)^2 + w_d\cdot (\epsilon_d - E_c)\Big]^{3/2}} ,
\end{equation}

\noindent where $E_c$ is the Coulomb potential at a separation of $R = r_0(A^{1/3} + 2^{1/3})$, $w_d = 0.37\cdot E_B$ is a parameter defining the width of the energy distribution (derived from a fit to literature data), and $E_B = 2.225$ MeV is the deuteron binding energy \cite{Wang_2021}.  $N_E$ is a normalization factor such that $\int_{E_n} P_{BU}(E_n) dE_n = 1$.  The $E_c$ factor was not included in the original Serber theory \cite{serber1947production}, however it accounts for the slight shift in the centroid energy observed in measurements, particularly for higher $Z$ targets, due to Coulomb repulsion of the incident deuteron.

The angle distribution from the Serber model is given by:

\begin{align}
P_{BU}(\theta) =& \ \  N_{\theta} \cdot \frac{\theta_0}{2\pi(\theta_0^2 + \theta^2)^{3/2}}, \nonumber \\
 \theta_0 =& \ \  0.72 \cdot \sqrt{\frac{E_B}{\epsilon_d - E_c}} \cdot \Big(1-\frac{\epsilon_d}{8m_dc^2}\Big) ,
\end{align}

\noindent where $m_d$ is the deuteron mass, and $N_{\theta}$ is again a normalization factor such that $\int_{\theta} P(\theta) d\theta = 1$.  


\begin{figure*}[!t]
    \sloppy
    \centering
    \subfloat{
        \centering
        \subfigimg[width=0.496\textwidth]{}{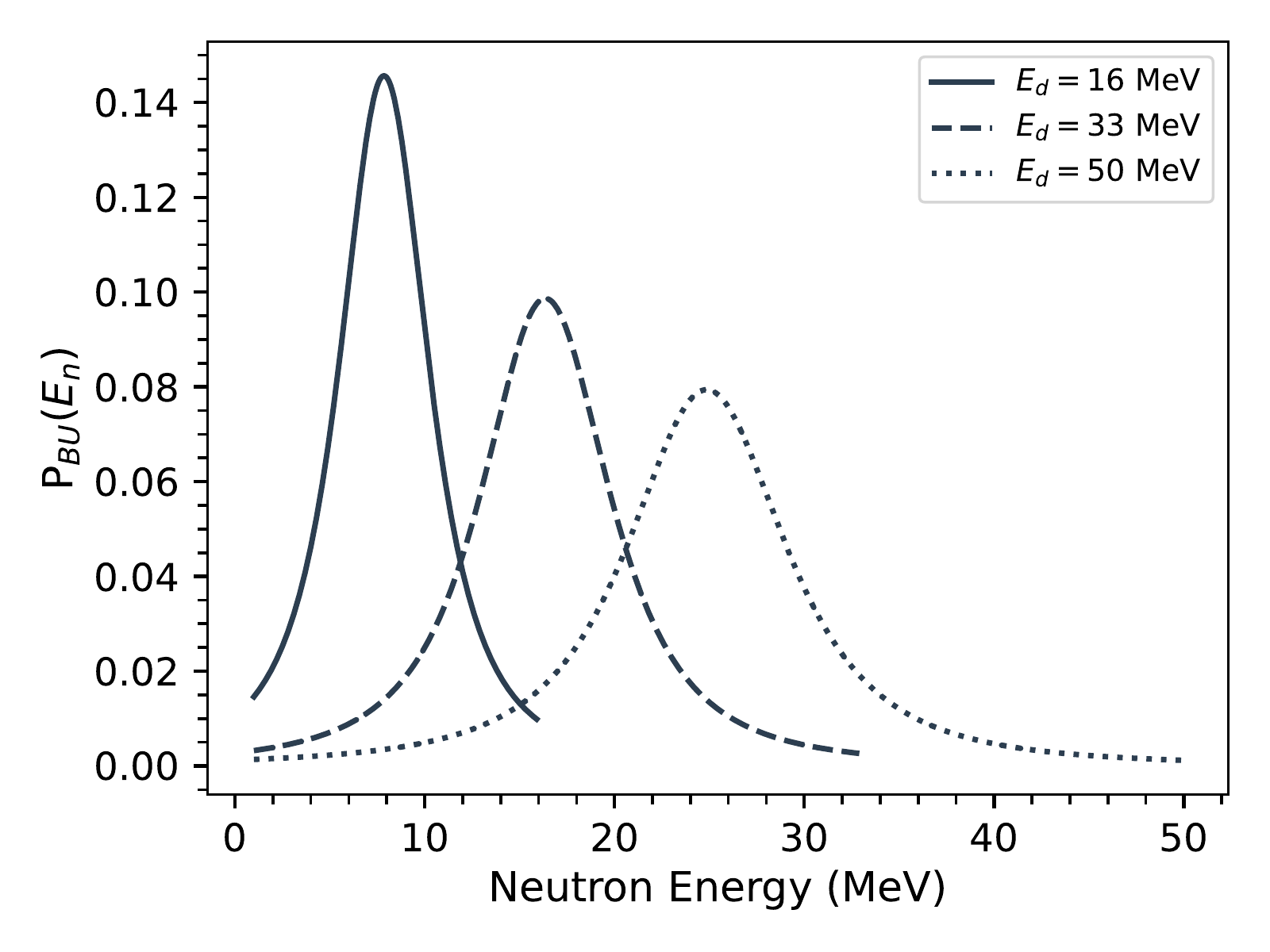}{50}
        \subfigimg[width=0.496\textwidth]{}{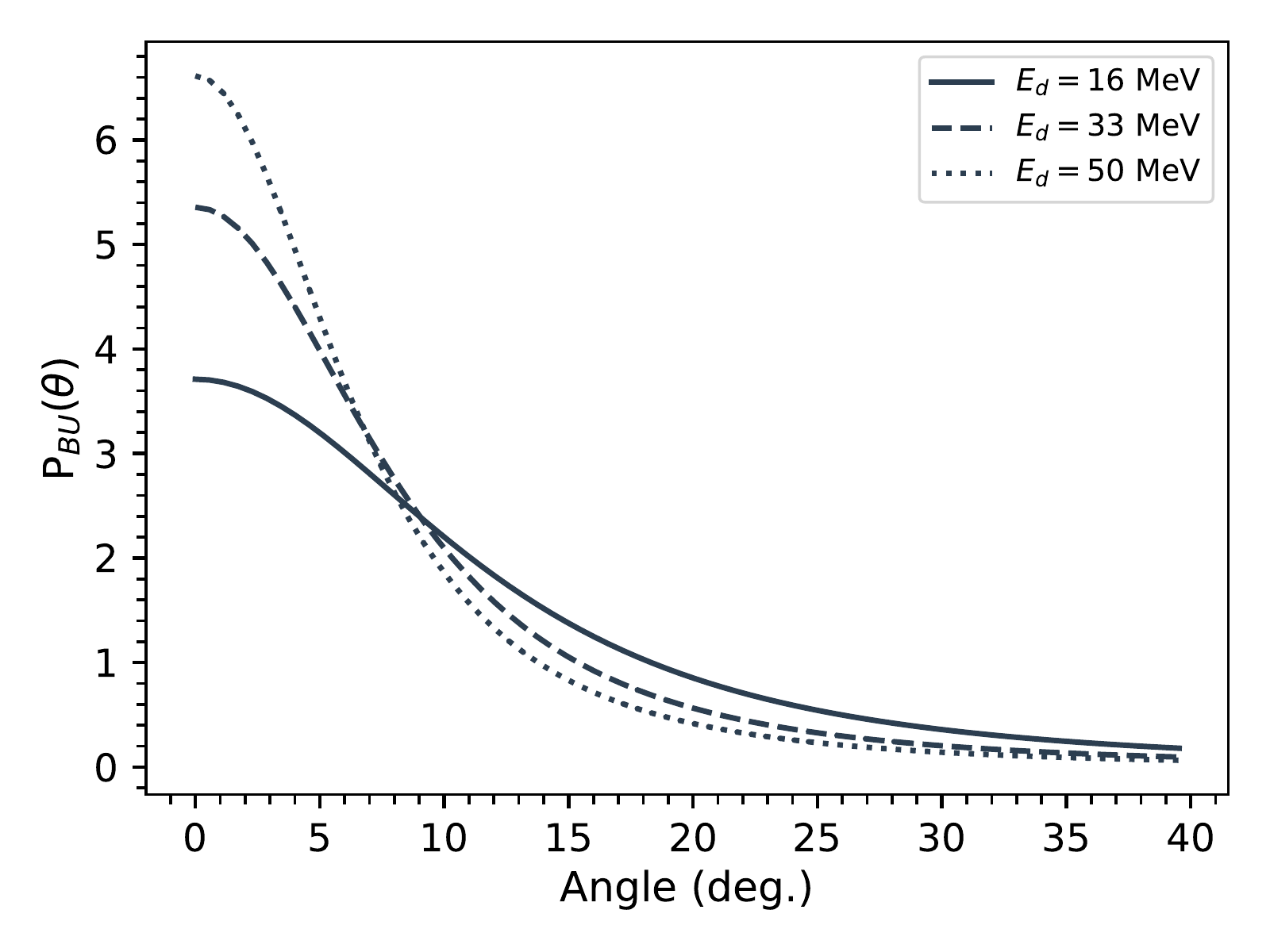}{50}
   \hspace{-10pt}}
    \caption{Energy (left) and angle (right) distributions of the outgoing neutrons from the deuteron breakup component, for various incident deuteron energies.}
   	\label{fig:bu_distributions}
\end{figure*}

Figure \ref{fig:bu_distributions} shows these distributions for the breakup reaction on a beryllium target.  Characteristic of the breakup process, the neutron energy distribution is centered at half the incident deuteron energy (minus the Coulomb potential), with a width that increases as the incident energy increases.  The angular distribution is very forward-focused, with the majority of breakup neutrons emerging below about 20$^{\circ}$.  This angular distribution narrows with increasing energy.

However, breakup is not the only reaction responsible for neutron production.  While the pre-equilibrium and compound nuclear reactions could be calculated using a standard nuclear model code such as TALYS \cite{TALYS} or EMPIRE \cite{HERMAN20072655}, for the purposes of a parametric study we have made use of semi-empirical models for these mechanisms, largely based on the work of Kalbach \cite{kalbach1988systematics}.  This was done so that the pre-equilibrium and compound reaction models could be adjusted to match experimental data, during the fitting procedure described in section \ref{literature_fitting}.

Similar to the breakup cross section, the compound and pre-equilibrium cross sections were assumed to have separable energy and angle distributions, according to:

\begin{equation}
\frac{d^2\sigma_x}{d\Omega dE_n} = \sigma_x(\epsilon_d)P_x(E_n)P_x(\theta) ,
\end{equation}

\noindent where the subscript $x$ refers to either the compound or pre-equilibrium process.

The total compound neutron production cross section was given the following empirical relation in terms of deuteron energy:

\begin{align}
\sigma_{CM}(\epsilon_d) = 80.6\cdot \Big[&\text{exp}\Big(-\frac{1}{2}\big(\frac{18-\epsilon_d}{14}\big)^2\Big) \nonumber \\
&+ \frac{0.3}{1+e^{(18-\epsilon_d)/7}} \Big] \text{ (mb)} .
\end{align}

The total pre-equilibrium neutron production cross section is described in a similar empirical relation according to:

\begin{equation}
\sigma_{PE}(\epsilon_d) =  \frac{34.2}{1+e^{(22-\epsilon_d)/6}} \text{ (mb)} .
\end{equation}

The systematics of the neutron angular distributions for compound and pre-equilibrium reactions have been thoroughly characterized in the semi-empirical formulations by Kalbach and Mann \cite{kalbach1981phenomenology}, and therefore we will adopt their parameterization according to the following relation:

\begin{equation}
P_{CM}(\theta) = \frac{a}{2\cdot \text{sinh}(a)} \cdot \Big( \text{exp}\big(a \cdot \text{cos}(\theta')\big) + \text{exp}\big(-a\cdot \text{cos}(\theta') \big) \Big) ,
\end{equation}

\noindent where $\theta'$ is the neutron emission angle from the center-of-mass frame, and $a$ is the ``little a" parameter as defined by Kalbach and Mann's 1981 systematics \cite{kalbach1981phenomenology}. This $a$ parameter is dependent on the incident particle energy, and both the target and projectile mass and atomic number.

\begin{figure*}
    \sloppy
    \centering
    \subfloat{
        \centering
        \subfigimg[width=0.496\textwidth]{}{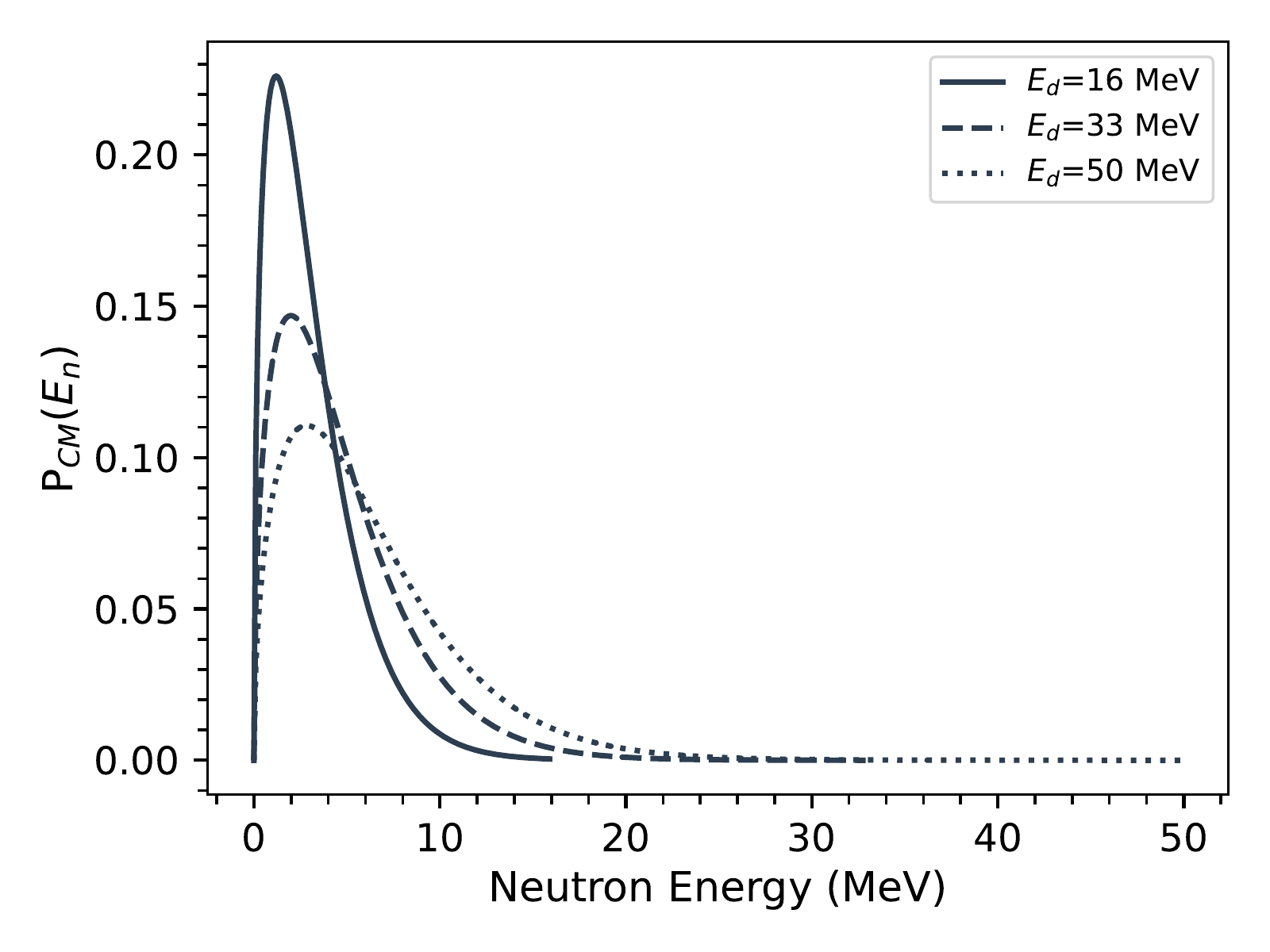}{50}
        \subfigimg[width=0.496\textwidth]{}{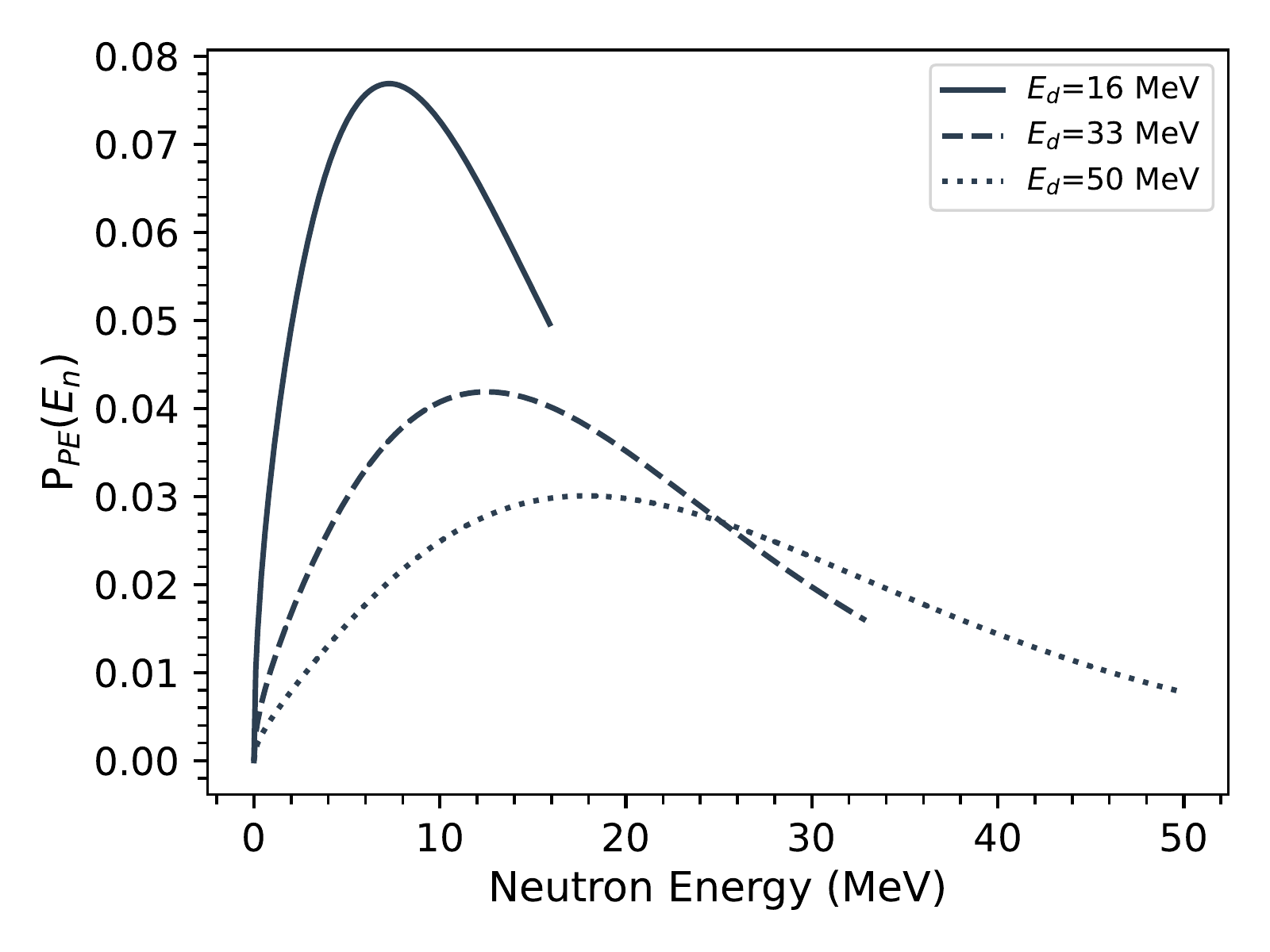}{50}
   \hspace{-10pt}}
    \caption{Energy distributions of compound (left) and pre-equilibrium (right) reaction components, determined using a modified moving-source parameterization.}
   	\label{fig:energy_distributions}
\end{figure*}

The pre-equilibrium angular distribution is given by a similar formula:

\begin{equation}
P_{PE}(\theta) = \frac{a}{\text{sinh}(a)} \cdot  \text{exp}\big(a \cdot \text{cos}(\theta')\big) .
\end{equation}

For the case of a beryllium target, it was found that the compound and pre-equilibrium angular distributions were somewhat more forward-focused than predicted by the Kalbach-Mann systematics, and thus the values of $a$ were modified, with the values $a_{CM} = 1.1 \cdot a$ and $a_{PE} = 1.8 \cdot a$ giving the best fit to literature data.

The neutron energy distributions for compound and pre-equilibrium reactions were parameterized using a Watt distribution, according to:

\begin{equation}
P_x(E_n) = N_x\cdot \text{sinh}\big(\sqrt{2\cdot E_n}\big) \cdot \text{exp}\big(\frac{-E_n}{kT_x} \big) ,
\end{equation}

\noindent where $kT_x$ is the nuclear temperature associated with each process.  The Watt distribution was found to reproduce the experimental results better than the Maxwell-Boltzmann distribution, which is intuitive given that a Watt distribution is characteristic of a moving source that emits particles with a Maxwell-Boltzmann distribution in the center-of-mass frame.  Once again, $N_x$ is a normalization factor such that $\int_0^{\epsilon_d} P_{CM,PE}(E_n) dE_n = 1$.

The energetic dependence of the nuclear emission temperature was loosely based on the parameterization from the moving-source model of Cronqvist \cite{cronqvist1990moving}, with:

\begin{equation}
kT_{CM} = 0.1 + 0.27\cdot \sqrt{\epsilon_d + Q} \text{ (MeV)} , 
\end{equation}

and:

\begin{equation}
kT_{PE} = 0.75 + 0.63\cdot \sqrt{\epsilon_d + Q} \text{ (MeV)} ,
\end{equation}

\noindent where $Q$ is 4.36 MeV for \ce{^{9}Be}(d,n).

Figure \ref{fig:energy_distributions} plots the energy distributions for compound and pre-equilibrium emission.  These give the expected trends that pre-equilibrium emission will result in a comparably higher average energy, and that the distribution spreads out as the deuteron energy increases.  For a more rigorous calculation of these distributions, one could make use of a Hauser-Feshbach calculation \cite{hauser1952inelastic} for the compound spectrum and an exciton-model calculation \cite{blann1971hybrid} for the pre-equilibrium distribution.  However this was not performed here for the sake of simplicity, which is justified by the fact that these reactions form a fairly minor contribution to the breakup spectrum.

It should be noted that the black-body temperature $kT$ is not an absolute temperature in any moving reference frame, and there will be an apparent temperature shift given by the relativistic correction:

\begin{equation}
T'(\theta', \epsilon_d) = \frac{T}{\gamma\cdot \big(1-\beta \cdot \text{cos}(\theta')\big)} ,
\end{equation}

\noindent where $\beta=\frac{v}{c}$ is the relativistic velocity for the compound system (in the lab frame), $\gamma$ is the Lorentz factor $(1-\beta^2)^{-1/2}$, and $\theta'$ is again the neutron emission angle in the center of mass frame.


\begin{figure*}[!t]
    \sloppy
    \centering
    \subfloat{
        \centering
        \subfigimg[width=0.496\textwidth]{}{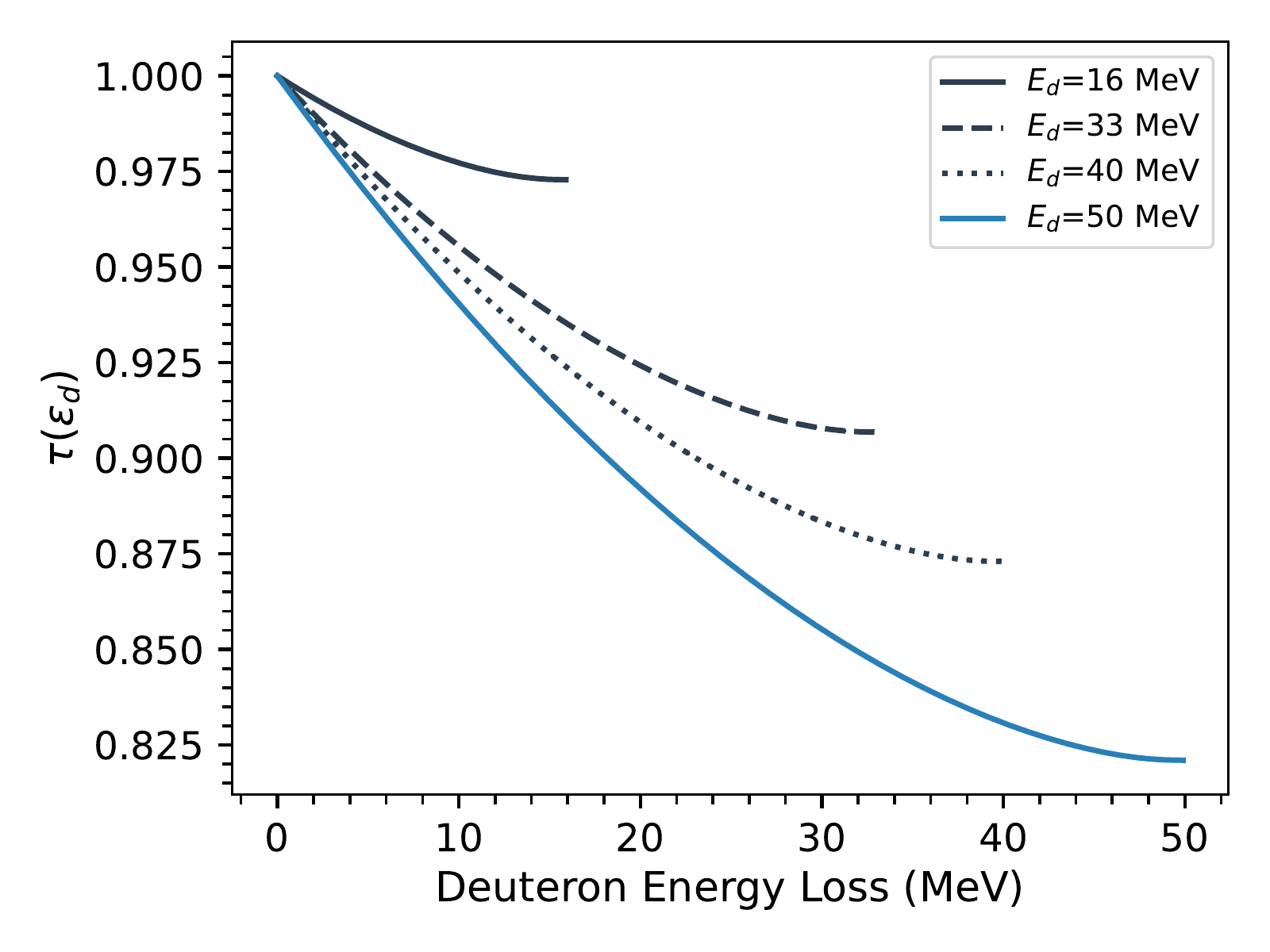}{50}
        \subfigimg[width=0.496\textwidth]{}{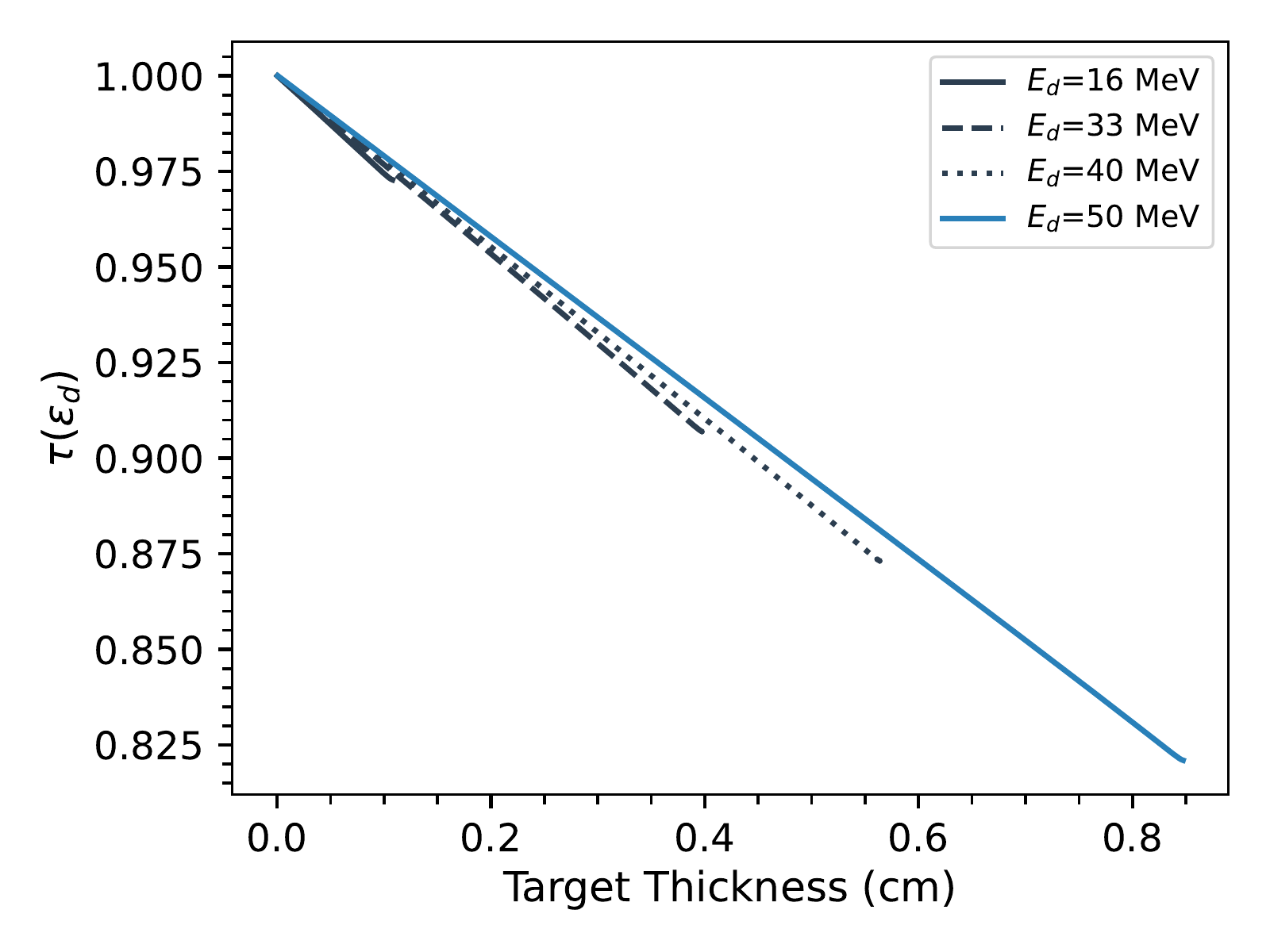}{50}
   \hspace{-10pt}}
    \caption{Deuteron transmission factor ($\tau$) as a function of the deuteron energy loss in the target (left) and the target thickness (right) for a range of incident energies.  Note the zero-suppressed y-axis (the minimum transmission at 50 MeV is about 82\%).}
   	\label{fig:tau_plots}
\end{figure*}

The consequence of this correction is a relatively small shift in the energy distribution, however it does induce a correlation between the outgoing neutron emission energy and angle.  For example, at $\epsilon_d=40$ MeV the average black-body temperature is reduced by approximately 7.5\% between 0$^{\circ}$ and 150$^{\circ}$: a small but meaningful shift, particularly at large angles where the compound contribution to the neutron spectrum is dominant.

All that remains in order to calculate thick target yields is an expression for the transmission parameter $\tau$.  Given the expected linear trajectory of deuterons in a material down to energies below the threshold for breakup, the transmission parameter can be approximated as the integration of the total reaction probability, according to:

\begin{equation}
\tau(E_d) = \text{exp}\Big( -\rho_N \int_{\epsilon_d}^{E_d}  \sigma_T(\epsilon')\big(\frac{d\epsilon'}{dx}\big)^{-1} d\epsilon' \Big) ,
\end{equation}

\noindent where $\rho_N$ is the number density of the target, $\sigma_T$ is the total reaction cross section (not just neutron producing reactions), $E_d$ is the original deuteron energy incident on the target, and $\frac{d\epsilon'}{dx}$ is the stopping power.  In this work the Anderson \& Ziegler formalism was used to determine the deuteron stopping powers \cite{ZIEGLER20101818}, however at the energies relevant to deuteron breakup the Bethe formula would have likely sufficed.

This total reaction cross section, $\sigma_T(\epsilon')$ can be interpolated from an evaluation such as TENDL, however for the sake of computational simplicity we have used the following relation:

\begin{equation}
\sigma_T(\epsilon') = \sigma_{BU}(\epsilon') + r_0^2 (A^{1/3} + 0.8)^2 \cdot \Big[c_1 \cdot e^{-\epsilon'/a_1} \cdot (1-e^{-\epsilon'/a_2}) \Big] ,
\end{equation}

\noindent where the parameters $c_1 = 5.643$, $a_1=131.3$ MeV, and $a_2=1.354$ MeV were derived from a least-squares fit to TALYS-1.9 calculations on Li, Be and C targets \cite{TALYS}.  In this case $\sigma_{BU}$ is the contribution from breakup only, i.e., not including compound or pre-equilibrium reactions.  

The transmission parameter $\tau$ can be seen in the plots of Fig. \ref{fig:tau_plots} as a function of both the target thickness and the deuteron energy loss in the target.  A significant feature of these plots is that even at higher incident energies, the deuteron beam is not highly attenuated -- approximately 20\% of a 50 MeV deuteron beam is attenuated in a thick Be target.  This leads to a general trend in the thick target deuteron breakup measurements: that the breakup spectrum for a given energy can be very nearly approximated by the integral of the breakup spectra from all energies below it.  For example, the neutron spectrum from 40 MeV deuterons should be approximately equal to the 30 MeV spectrum, plus the additional flux of the deuterons ranging from 40 down to 30 MeV.  This highlights inconsistencies in the literature very clearly.  For example, if the reported integral yields at 30 MeV (deuteron energy) are higher than another measurement at 40 MeV, one of the two measurements must be incorrect.

\subsection{Literature Data Selection and Fitting}\label{literature_fitting}

The emphasis of this work is on reliably calculating double-differential thick target neutron yields on beryllium.  Lithium and carbon targets would have also been appropriate, as the stripping reaction mechanism should be the dominant contribution to breakup for low-Z nuclides; however, there are fewer experimental data on these targets.  Instead, while the data on lithium and carbon (and copper) targets were not included in the fitting procedure, they were used to evaluate the predictive capability of the model to be extrapolated outside the range of where it was fitted.  It may be of value to perform the same dedicated fitting procedure on lithium targets, as these are interesting for applications as they produce an approximately 30\% higher neutron yield per unit of beam current, due to the increased deuteron range in lithium.  Furthermore, there are existing high-power liquid lithium targets that are being used for neutron generation for astrophysical applications \cite{paul2019reactions}.  However, this analysis will be left for a future study.

In total, the measurements of five authors were found with thick target neutron yield data on beryllium targets in the relevant energy range.  The most recent, Harrig \textit{et al.}, performed very precise measurements using the double time-of-flight technique at 16 MeV, at forward angles \cite{harrig2018neutron}.  The use of this method allowed Harrig to measure significantly lower neutron energies than other authors, which all used the (single) time-of-flight method.  The works of Saltmarsh \cite{saltmarsh1977characteristics} and Meulders \cite{meulders1975intensity} were quite valuable, because they contained measurements at multiple angles.  This is important for assessing the contributions from compound and pre-equilibrium reactions, as these mechanisms will be observed relatively free of breakup neutrons at large angles.  However, the 33 MeV data of Meulders were not used in the fitting procedure, as the integral yield of that particular measurement was extraneously high compared to the rest of the available literature data.  

Also, the Schweimer data \cite{schweimer1967fast} were interesting because they contained a measurement at 54 MeV, which would have extended the range of incident deuteron energies, but had to be excluded because the 40 MeV Schweimer measurement was about 40\% lower in magnitude than other literature data, and it was unclear if a normalization error also affected the 54 MeV measurement.  Weaver \cite{weaver1973neutron} had multi-angle measurements at 16 MeV, however because the present study does not include direct reactions, these low energy points were highly discrepant.  Also, there was an approximately 25\% discrepancy between Weaver and the Harrig/Meulders measurements at 16 MeV.  Therefore the Weaver and Schweimer data were excluded from the fitting; however, we will show a comparison of these data to the resultant model in the next section.

Six of the parameters from the proposed model were selected to be adjusted in the fitting procedure.  These six parameters were: the scaling parameters for the magnitudes of the breakup, compound and pre-equilibrium reaction components ($\sigma_{BU}$, $\sigma_{CM}$ and $\sigma_{PE}$), a parameter for the widths of the breakup energy ($w_d$) and angle distributions ($\theta_0$), as well as a slope parameter $\eta_{BU}$ for the breakup total cross section.  These parameters were fit with a least-squares method, making use of six scaling constants $c_i$ according to:

\begin{align}
\frac{d^2\sigma (c_1,\ldots ,c_6)}{d\Omega dE_n} =& \ \ c_1\cdot \frac{d^2\sigma_{BU}(c_2\cdot \eta_{BU}, c_3\cdot w_d, c_4\cdot \theta_0)}{d\Omega dE_n} \nonumber \\
 +& c_5\cdot \frac{d^2\sigma_{CM}}{d\Omega dE_n} + c_6\cdot \frac{d^2\sigma_{PE}}{d\Omega dE_n} \label{eq:fitting}
\end{align}

The results of this fit were incorporated into the previously described model with our ``recommended" parameters, which was re-normalized such that all $c_i=1$.  This 6-parameter function will be useful for the neutron activation analysis measurements performed in section \ref{measurements_section}, as it can be used to extract the measured spectrum without the issue of solving an underdetermined system, typical of spectral unfolding techniques.

\begin{figure}[htb]
\centering
\includegraphics[width=0.99\linewidth]{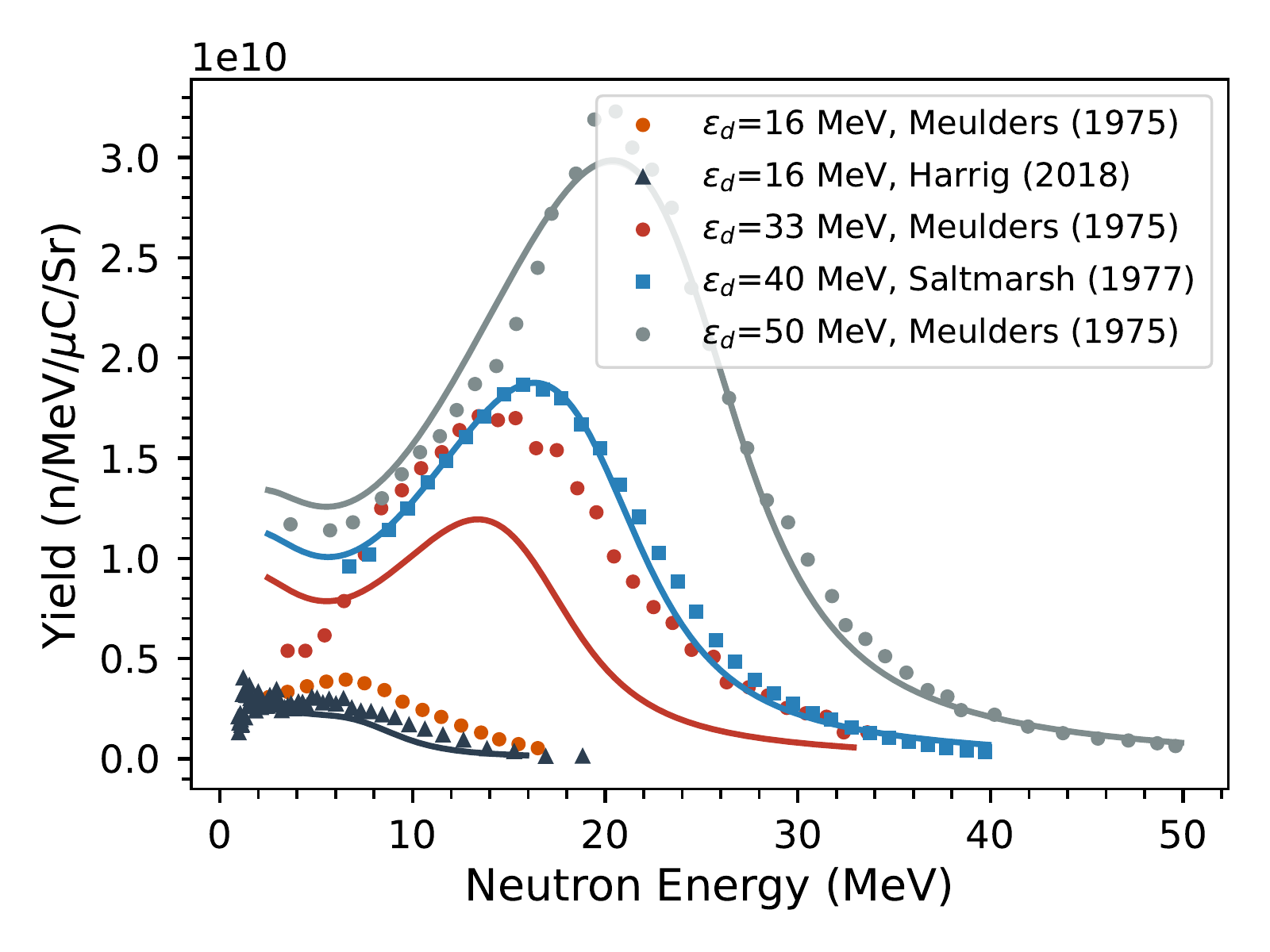}
\caption{Results of the fit to literature data measured at $\theta=0^{\circ}$, from $\epsilon_d=16$--50 MeV.  Data from Meulders (1975), Saltmarsh (1977) and Harrig (2018) \cite{meulders1975intensity, saltmarsh1977characteristics, harrig2018neutron}.  Note that the 33 MeV Meulders data were not included in the fit procedure.
}
\label{fig:lit_0deg}
\end{figure}

\begin{figure*}[!t]
    \sloppy
    \centering
    \subfloat{
        \centering
        \subfigimg[width=0.496\textwidth]{}{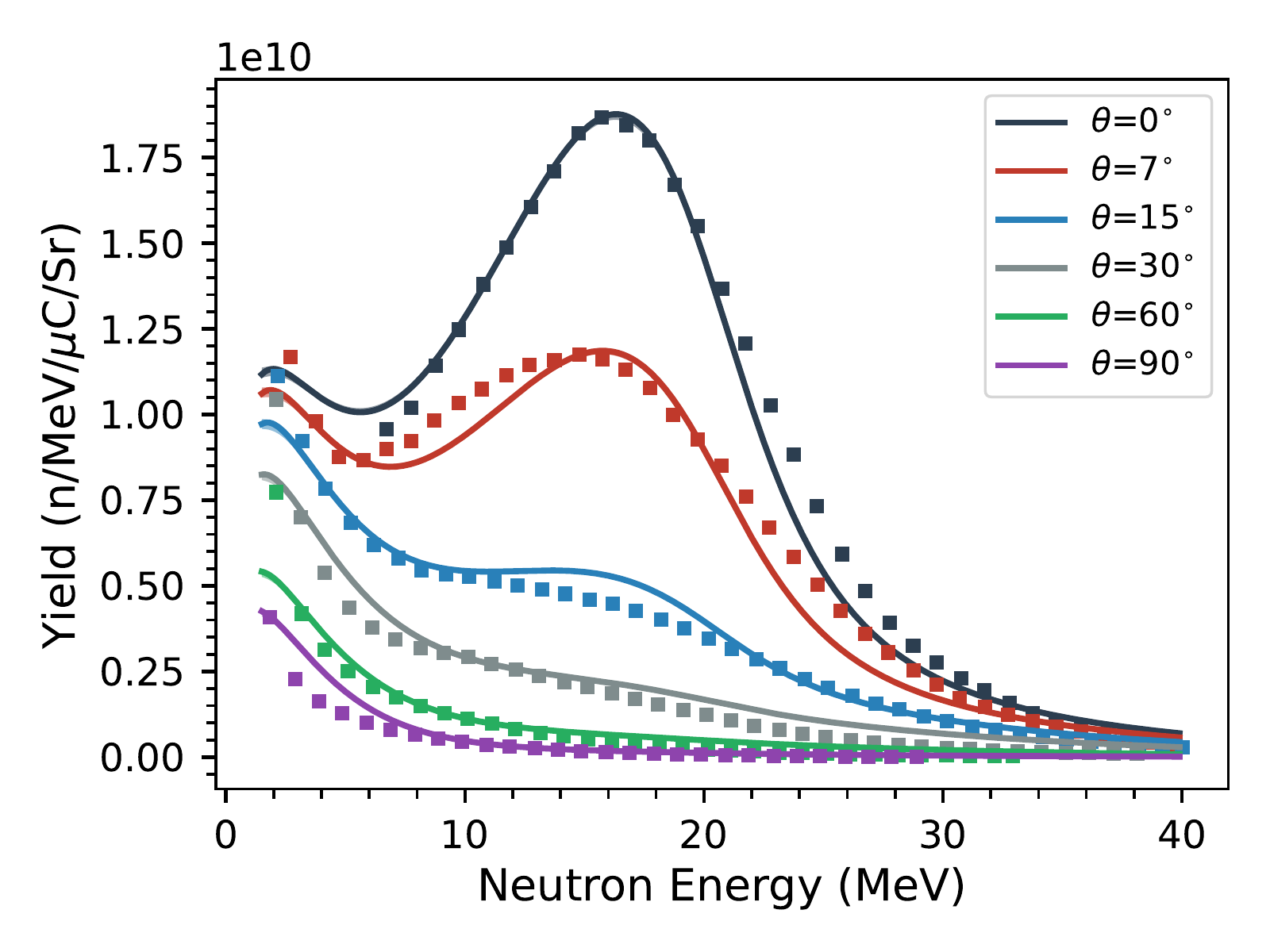}{50}
        \subfigimg[width=0.496\textwidth]{}{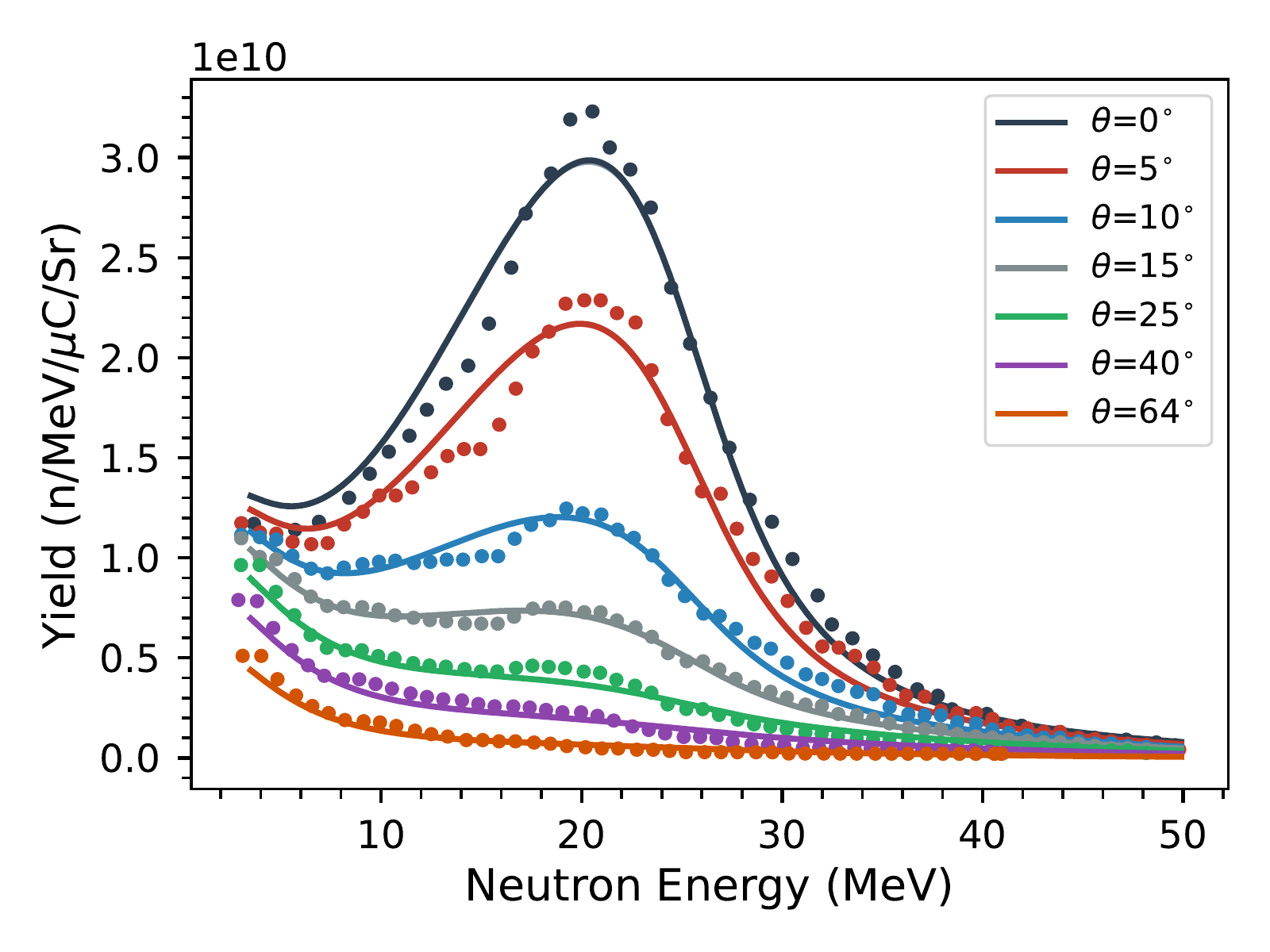}{50}
   \hspace{-10pt}}
    \caption{(left) Results of the fit to measurements by Saltmarsh (1977) at $\epsilon_d=40$ MeV from $\theta=0^{\circ}$ to $\theta=90^{\circ}$. (right) Results of the fit to measurements by Meulders (1975) at $\epsilon_d=50$ MeV from $\theta=0^{\circ}$ to $\theta=64^{\circ}$ \cite{saltmarsh1977characteristics, meulders1975intensity}.   
    }
   	\label{fig:lit_angles}
\end{figure*}

Determining the values of $c_i$ which best fit the literature data was performed by computing the thick target yields using Eq. (\ref{eq:fitting}), and iteratively adjusting $c_i$ to improve their fit to the experimental yields.  It is worth noting that this fitting procedure is performed once for all literature data, at all energies and angles together, rather than for a single spectrum at a time.  This reduces the likelihood of ``overfitting", improving the predictive power of this model.

The results of this fitting procedure to the selected literature data can be seen in Fig. \ref{fig:lit_0deg} at forward angles, and in Fig. \ref{fig:lit_angles} at $\epsilon_d = 40$ and 50 MeV for multiple outgoing angles.  The relative uncertainty for each fitted parameter $c_i$ is given in Table \ref{table:lit_uncertainty}.

\begin{table}[htb]
\centering
\begin{tabular}{clc}
\hline \hline
$i$ \ \ & Parameter Description \ \ & Uncertainty in $c_i$ (\%) \ \ \Tstrut \Bstrut \\
\hline
\Tstrut 1 & $\sigma_{BU}$ Magnitude & 2.2  \\ 
2 & $\sigma_{BU}$ Slope &  7.4  \\ 
3 & $P_{BU}(E_n)$ Width & 3.8  \\ 
4 & $P_{BU}(\theta)$ Width & 1.3 \\
5 & $\sigma_{CM}$ Magnitude & 2.0  \\
6 & $\sigma_{PE}$ Magnitude & \hspace{0.08em} 3.8 \Bstrut \\
\hline \hline
\end{tabular} 
\caption{Relative uncertainty for each of the parameters, $c_i$, determined by the fit to literature data.  Recall that the previously described model has been re-normalized such that all $c_i=1$.
}
\label{table:lit_uncertainty}
\end{table}

In general, the agreement is quite good. The results seem to indicate that the Meulders 33 MeV measurement is about 40\% too high, rather than the Saltmarsh 40 MeV measurement being too low.   There is a slight disagreement between the suggested width of the breakup peak in the 40 MeV data by Saltmarsh, which seems to show a wider peak, and in the 50 MeV Meulders data, which has a narrower peak.  Also, because this method neglects direct reactions (for being outside the focus on isotope production applications), there is an underprediction of the 16 MeV Harrig dataset.  There is also a subtle disagreement in the low energy compound spectrum, which could be related to issues in determining the time-of-flight detection efficiency at low energies.

\subsection{Model Validation}

\begin{figure*}[!h]
    \sloppy
    \centering
    \subfloat{
        \centering
        \subfigimg[width=0.496\textwidth]{}{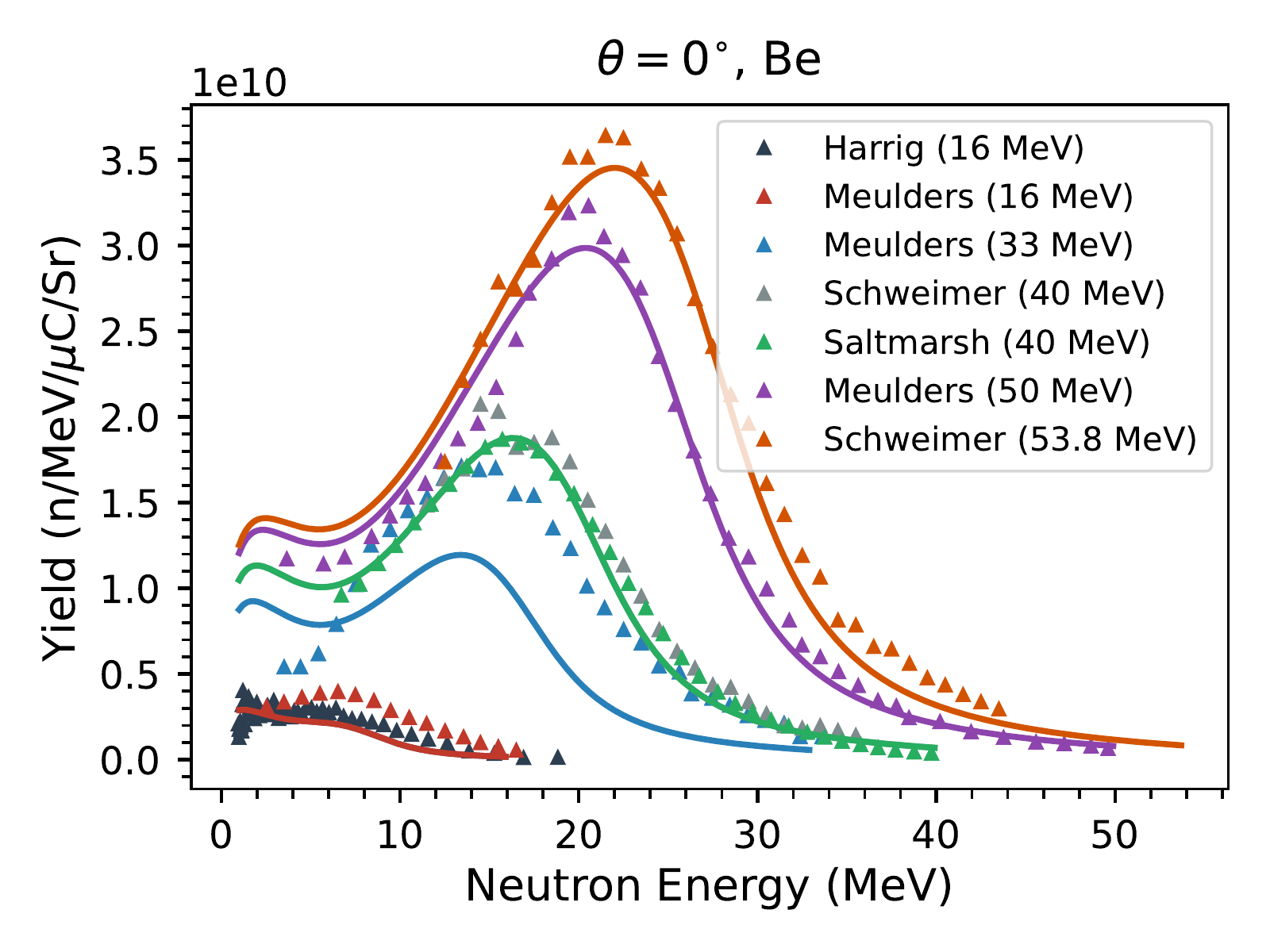}{50}
        \subfigimg[width=0.496\textwidth]{}{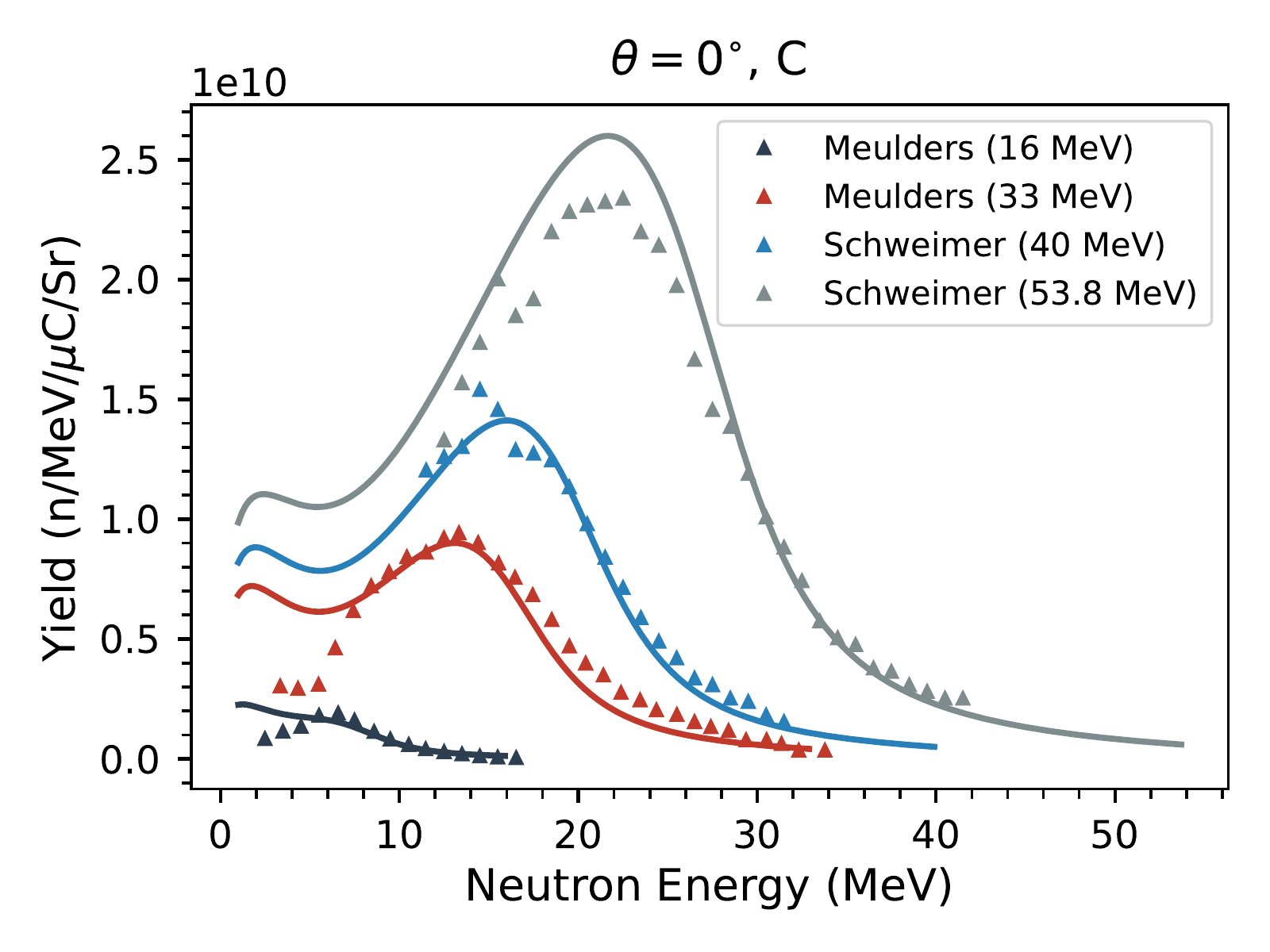}{50}
   \hspace{-10pt}}
    \\
    \subfloat{
        \centering
        \subfigimg[width=0.496\textwidth]{}{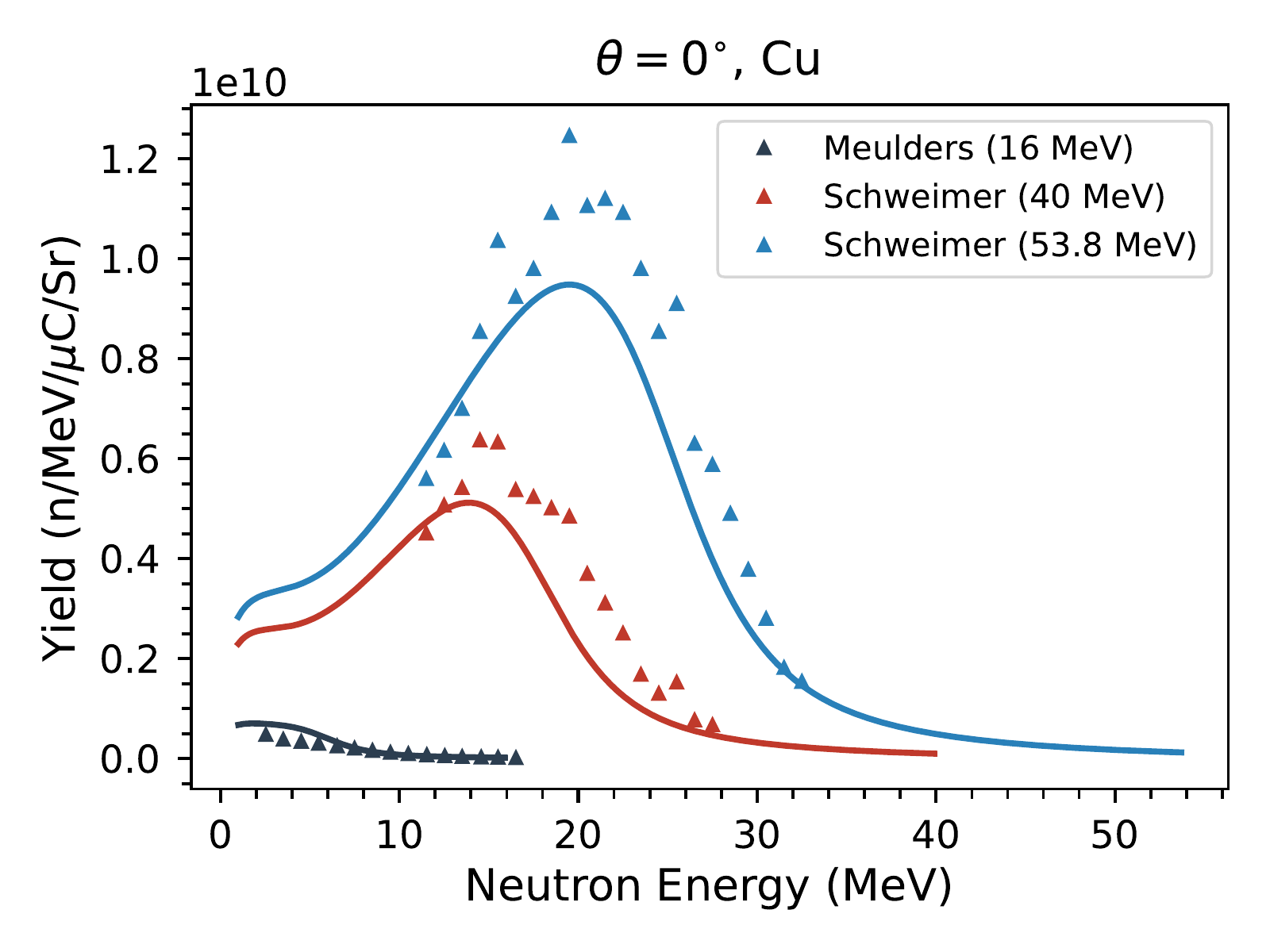}{50}
        \subfigimg[width=0.496\textwidth]{}{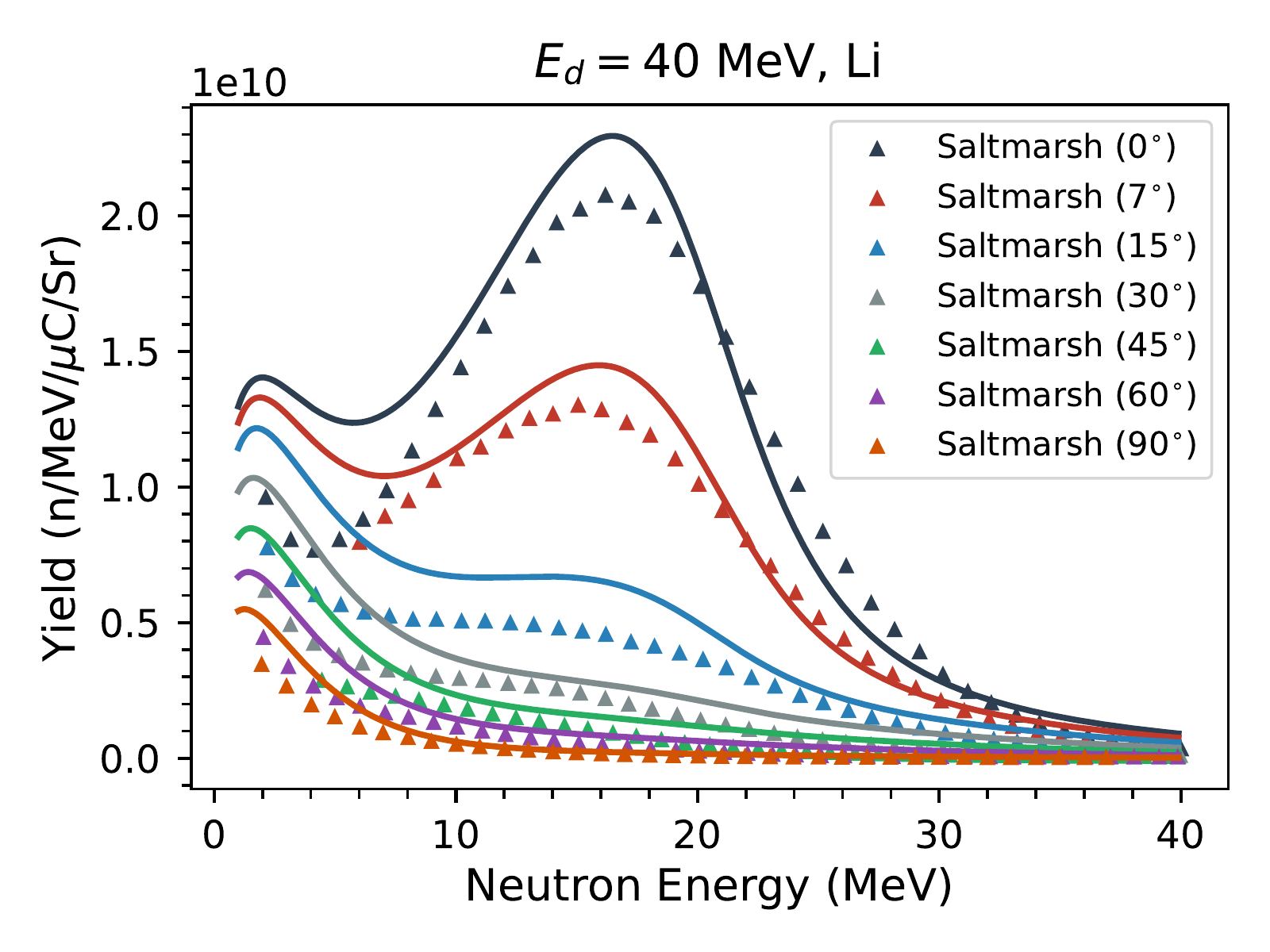}{50}
   \hspace{-10pt}}
    \\
    \subfloat{
        \centering
        \subfigimg[width=0.496\textwidth]{}{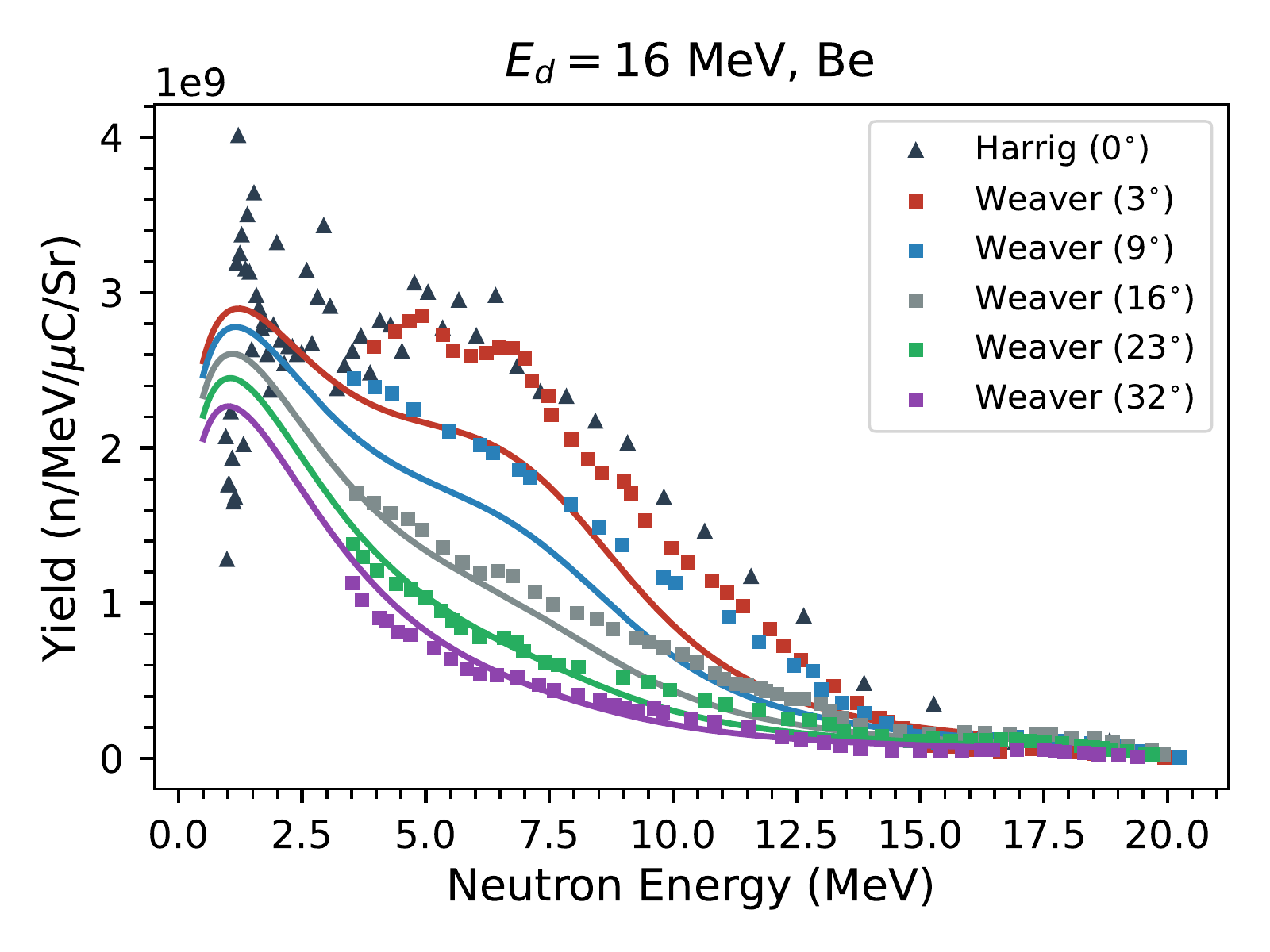}{50}
        \subfigimg[width=0.496\textwidth]{}{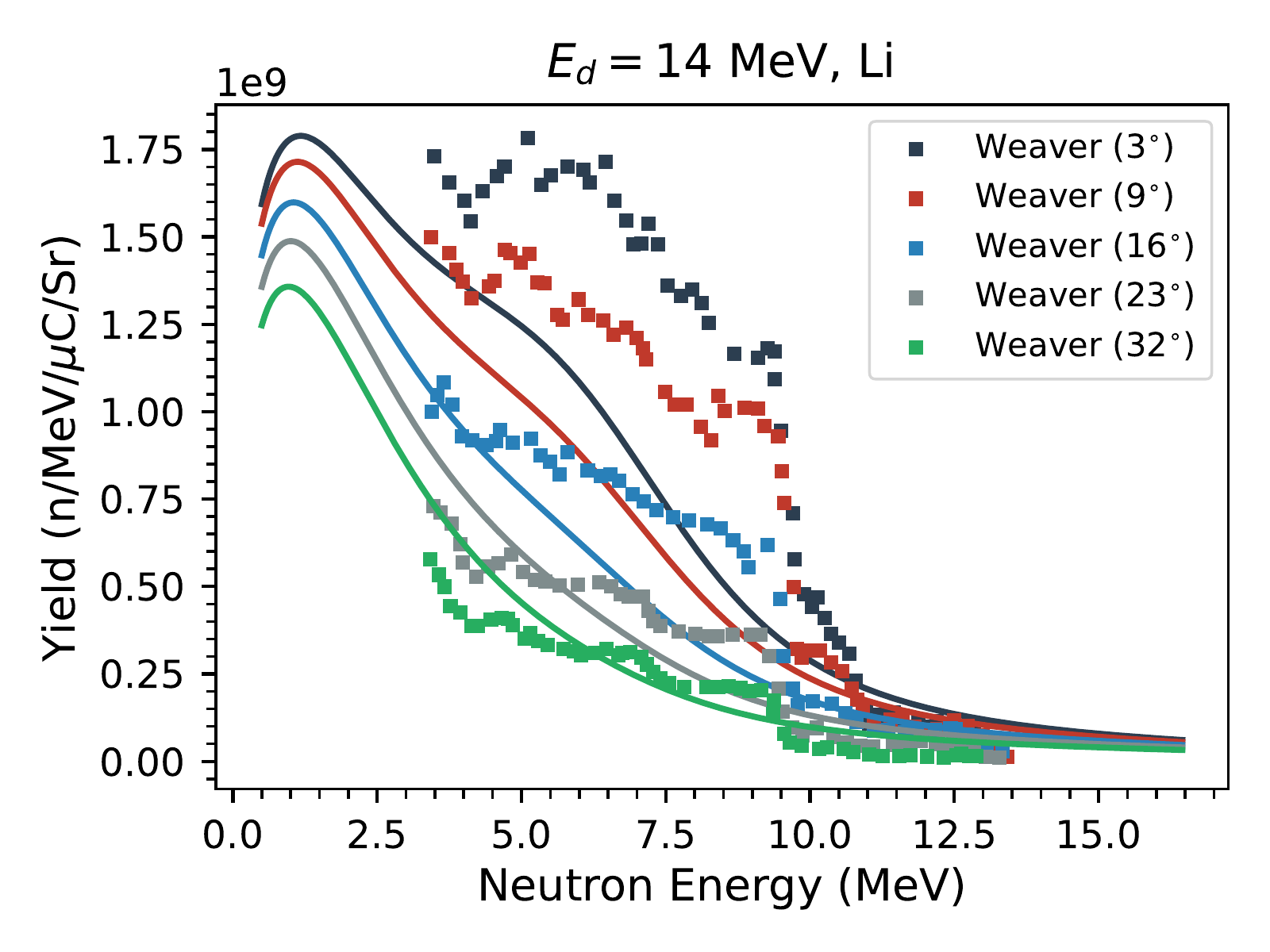}{50}
   \hspace{-10pt}}
    \caption{Comparison of modeled neutron yields to literature measurements \cite{meulders1975intensity, saltmarsh1977characteristics, harrig2018neutron, schweimer1967fast, weaver1973neutron}. The first three plots show $\theta=0^{\circ}$ measurements on Be, C and Cu.  The following three plots show measurements at multiple angles, on Li and Be.    
    }
    \label{fig:validation}
\end{figure*}

The collected data that was not used for fitting the breakup model was instead used in a validation procedure, to estimate how well the model extrapolated to other materials and deuteron energies.  The results of this validation can be seen in Fig. \ref{fig:validation}.  Because of the previously mentioned discrepancies, the plotted Schweimer data were multiplied by a factor of 1.4, in order to align with Saltmarsh at 40 MeV, and the Weaver data were multiplied by 0.8, in order to match the Harrig data at 16 MeV.  

The results of this validation are encouraging.  The model extrapolates well to the \emph{corrected} 54 MeV Schweimer data. This builds confidence in the model at higher energies, an important consideration for isotope production applications.  The modeled yields on lithium and carbon targets are systematically higher than the data, however this seems mostly attributable to the compound component of the cross sections.  This is quite evident at low energies, and at large angles, where the compound mechanism is dominant.  This is not surprising, as the model used for the compound (and pre-equilibrium) contributions to the spectrum was specifically optimized to match beryllium data, rather than being based on a more broadly applicable theory.  Interestingly, our hybrid breakup model extrapolates well to copper ($Z=29$), despite the omission of elastic breakup mechanisms (as it is based on the Serber proton-stripping theory).  At low energy, the model largely fails to reproduce the double-differential data from Weaver, except at large angles where we can assume most of the neutrons are from compound reactions.  This is likely due to the absence of direct reaction modeling, which is something that should be addressed in a future study.

\section{Measurements of Neutron Yields from Deuteron Breakup on a Thick Beryllium Target}\label{measurements_section}

As described in the previous section, there are a number of discrepancies between existing literature measurements that motivate further experimental clarification.  In this work, we present the results of two separate irradiations to measure the neutron yields from thick target deuteron breakup on beryllium.  Because the discrepancy surrounds the magnitudes of the 33 MeV Meulders dataset and the 40 MeV work of Saltmarsh, these were the two deuteron energies selected for these irradiations.

The first irradiation at 33 MeV consisted solely of a foil activation experiment, in which the fitting procedure employed in Eq. (\ref{eq:fitting}) was used to extract the observed spectrum from the foil data.  At 40 MeV, this was repeated, in addition to a time-of-flight measurement.  Each of these experiments consisted of measurements at multiple angles, such that the angular dependence of breakup, as well as the relative contributions from compound reactions and pre-equilibrium, could be understood.  In the foil activation experiments, zinc and titanium foils were also co-irradiated to quantify the production of certain medically-relevant radionuclides.

\subsection{Facility Overview}

Both irradiations were performed at the Lawrence Berkeley National Laboratory's 88-Inch Cyclotron \cite{One_Stop_Shop}.  The 88-Inch Cyclotron is a variable-beam, variable-energy K=140 isochronous cyclotron, capable of producing deuteron beams with energies up to 55 MeV.  The cyclotron facility has a number of shielded experimental ``caves", for various applications.  Both the 33 and 40 MeV irradiations were performed in Cave 0, considered the ``high level" cave due to the extensive radiation shielding enclosing it.  Critically for the time-of-flight experiment, this cave is well-shielded from neutrons produced by the deuteron beam scraping the extraction deflectors, which may produce neutrons that would interfere with the time-of-flight signals.  

One important consideration for the location was in the positioning of detectors for the time-of-flight experiment.  Due to the size of the cave, the EJ-309 neutron scintillators \cite{brown2018proton} were placed between 1.2--2.3 m from the breakup target, at angles ranging from 0--90$^{\circ}$ relative to the incoming beam axis.  While these path-lengths may be somewhat short, they still proved acceptable based on the energies being measured, the temporal resolution of the electronics, and the ambiguity attributable to neutron productions from temporally-adjacent beam bunches from the cyclotron (e.g., wrap-around).

Following irradiation, the neutron monitor foils were transferred to a separate counting room within the cyclotron laboratory.  These foils were counted with an ORTEC GMX Series (model GMX-50220-S) High-Purity Germanium (HPGe) detector.  For these irradiations, the detector energy and efficiency calibrations were determined with the use of the following NIST traceable standard calibration sources: \ce{^{152}Eu}, \ce{^{133}Ba}, \ce{^{137}Cs}, \ce{^{60}Co} and \ce{^{57}Co}.

\subsection{Activation Analysis}

Neutron activation analysis is a standard technique in one can measure the neutron flux over a specific energy range incident on a foil, by measuring the residual activation products induced via nuclear reactions.  The neutron spectrum is unfolded starting from the well known thin-target activation equation:

\begin{equation}
R = n \langle \sigma \rangle \langle \phi \rangle ,
\label{eq:activation}
\end{equation}

\noindent where $n$ is the number of target atoms, $R$ is the production rate, determined by the measured product activity, $\langle \sigma \rangle$ is the flux-averaged cross section given by:

\begin{equation}
\langle \sigma \rangle = \frac{\int_{E_n} \sigma(E_n) \phi(E_n) dE_n)}{\langle \phi \rangle} ,
\end{equation}

\noindent and $\langle \phi \rangle = \int_{E_n} \phi(E_n) dE_n$ is the scalar neutron flux.  Generally in order to solve this for  $\langle \phi \rangle$, a relative energy dependence of the flux must be assumed.  However for the purpose of this study, it is the shape of this flux that we are trying to determine.  In order to gain insight into the energy dependence, we measured multiple reaction channels in the neutron monitor foils, which each have different energy dependencies, making them sensitive to distinct energy regions.  In this way, we can essentially find the scalar flux in a small energy window, and then use this to gain information regarding the overall shape in an iterative manner.

\begin{figure}[htb]
\centering
\includegraphics[width=0.99\linewidth]{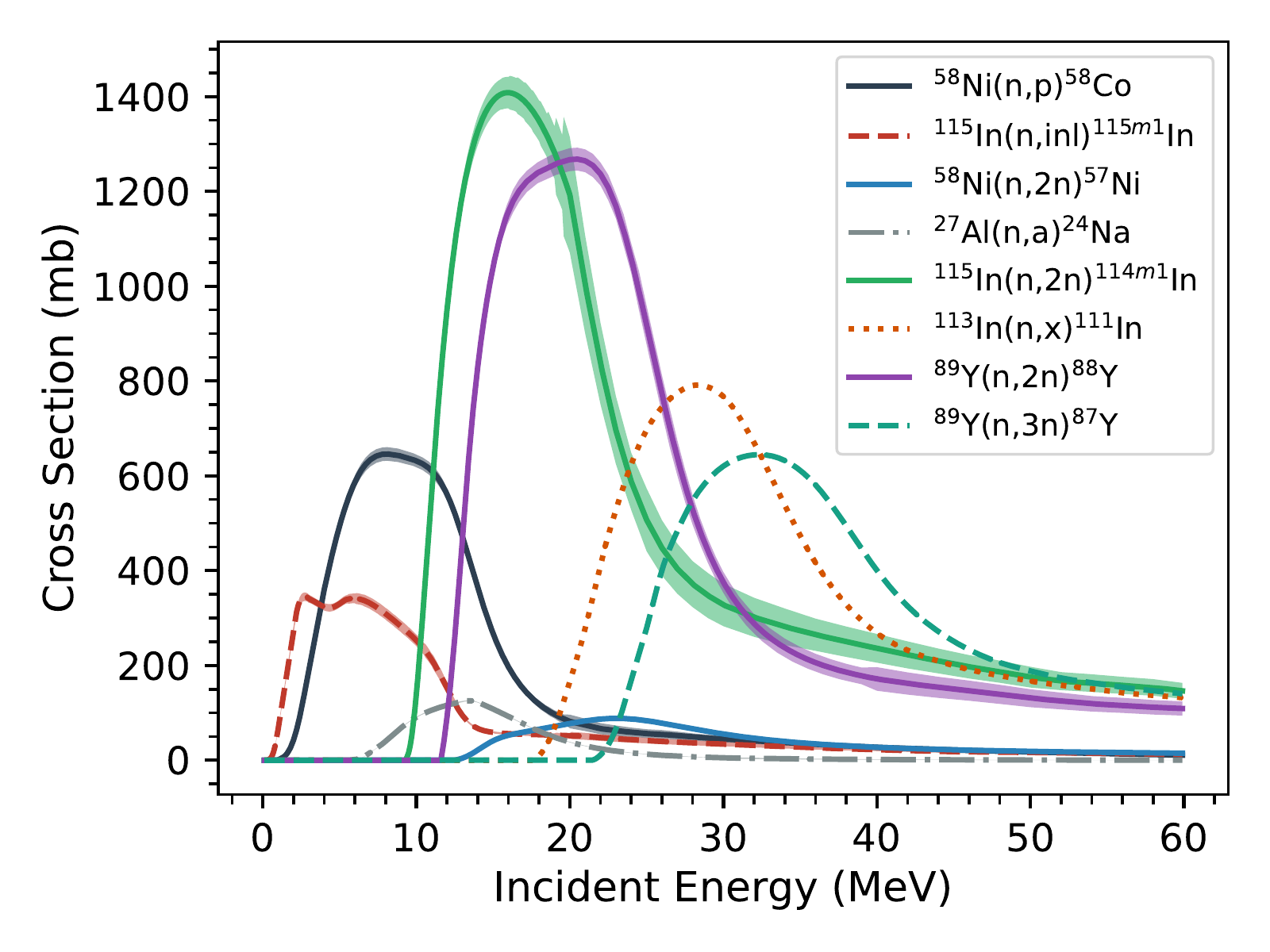}
\caption{Evaluated cross sections for some of the monitor reactions used in the foil activation experiment. 
}
\label{fig:monitor_reaction_xs}
\end{figure}

Figure \ref{fig:monitor_reaction_xs} demonstrates this with some of the monitor reactions that were used in this study.  It is clear that, for example, \ce{^{115}In}(n,n')\ce{^{115m}In} and \ce{^{89}Y}(n,3n)\ce{^{87}Y} have very little overlap, and are sensitive to neutrons in the 1--10 and 25--35 MeV energy ranges, respectively.  Therefore the relative activation rates of the two channels should provide the relative magnitude of the flux in each energy region, and consequently some idea about the flux shape.

If we insert our parameterized breakup model into the above activation equation, the predicted production rate is given by:

\begin{equation}
R_{pred,i}(\theta) = \frac{n\cdot I_{d}}{A} \int_{\hat{\Omega}} \int_{E_n=0}^{E_d} \sigma(E_n) \frac{d^2Y(E_n, \theta)}{d\Omega dE_n} dE_n d\Omega ,
\end{equation}

\noindent where $I_d$ is the deuteron beam current (in units of {\textmu}A), $A$ is the area of our monitor foil, and the double differential neutron yield $\frac{d^2Y(E_n, \theta)}{d\Omega dE_n}$, calculated using our hybrid breakup model, can be adjusted to match the measured production rates by varying the parameters ($c_1, \ldots, c_6$) in Eq. (\ref{eq:fitting}).  In addition to using multiple monitor foils, we can measure these production rates at multiple angles to improve the fit of the terms in Eq. (\ref{eq:fitting}) related to the breakup angular distribution, as well as the compound and the pre-equilibrium contributions to the spectrum, as these will be more sensitive at larger angles.

By comparing the predicted production rates for each of the $i$ reaction channels to those we measure through activation, $R_{meas,i}$, the optimum parameters $c_i$ can be fit by minimizing:

\begin{equation}
\chi^2 = \sum_i \frac{(R_{meas,i}-R_{pred,i})^2}{\sigma_{meas,i}^2} .
\end{equation}

This will yield the spectrum, in terms of our hybrid breakup model, which best reproduces the experimental reaction rates.

In order to produce the best results using this method, multiple activation foils were chosen that had well-characterized cross sections; either reactions that were evaluated in the IRDFF-II library \cite{trkov2020irdff} or where the TENDL-2015 evaluation gave good agreement with data from EXFOR \cite{rochman2017tendl}.  For the 33 MeV irradiation, nickel, indium, zirconium and aluminum monitor foils were used.  At 40 MeV, yttrium foils were also included in the foil packets.  Additionally, titanium foils were included in the 40 MeV irradiation, and zinc foils were included in both, in order to measure production cross sections for select radionuclides of potential interest for medical applications.  The specific monitor reactions used and libraries from which the cross section data were retrieved are given in Table \ref{table:activation_reactions}.

\begin{table}
\centering
\begin{tabular}{cc}
\hline \hline
Reaction & Library \Tstrut \Bstrut \\
\hline
\Tstrut  \ce{^{27}Al}(n,x)\ce{^{24}Na} & IRDFF-II  \\ 
\ce{^{nat}Ni}(n,x)\ce{^{58}Co} & IRDFF-II  \\ 
\ce{^{nat}Ni}(n,x)\ce{^{57}Ni} & IRDFF-II  \\ 
\ce{^{58}Ni}(n,x)\ce{^{57}Co} & TENDL-2015 \\
\ce{^{89}Y}(n,2n)\ce{^{88}Y} & IRDFF-II  \\
\ce{^{89}Y}(n,3n)\ce{^{87g}Y} & TENDL-2015  \\
\ce{^{89}Y}(n,3n)\ce{^{89m}Y} & TENDL-2015 \\
\ce{^{nat}Zr}(n,x)\ce{^{89}Zr} & IRDFF-II \\
\ce{^{113}In}(n,3n)\ce{^{111}In} & TENDL-2015 \\
\ce{^{115}In}(n,2n)\ce{^{114m1}In} & IRDFF-II \\
\ce{^{115}In}(n,n')\ce{^{115m}In} & \hspace{0.08em} IRDFF-II  \Bstrut \\
\hline \hline
\end{tabular} 
\caption{Neutron monitor reactions used for the foil activation spectral reconstruction.  Note that the reactions based on \ce{^{89}Y} targets only applied to the 40 MeV irradiation.  Cross section data were retrieved from IRDFF-II and TENDL-2015 \cite{trkov2020irdff, rochman2017tendl}.
}
\label{table:activation_reactions}
\end{table}

All foils were purchased from Goodfellow Corporation (Coraopolis, PA 15108, USA) and were of either 99 or 99.99\% purity (metals basis) with natural isotopic abundances.  All foils were cut to 1 cm diameter disks, and were cold rolled to 0.5 mm thickness, with the exception of the nickel foils which were cut to 6.3 mm diameter, of 1 mm thickness.  Each foil was cleaned using isopropyl alcohol, and then its mass was measured using a milligram balance (after drying).  One of each foil was then placed into a ``foil packet", which was sealed using thin pieces of Kapton polyimide tape.  Kapton with an acrylic based adhesive was specifically chosen to avoid contaminating \ce{^{28}Si}(n,x)\ce{^{24}Na} reactions from silicone based adhesives, which may have interfered with the \ce{^{27}Al}(n,$\alpha$) monitor channel \cite{VOYLES201853}, although the extent to which this poses a problem for neutron activation is likely minimal.

\begin{figure}[htb]
\centering
\includegraphics[width=0.7\linewidth]{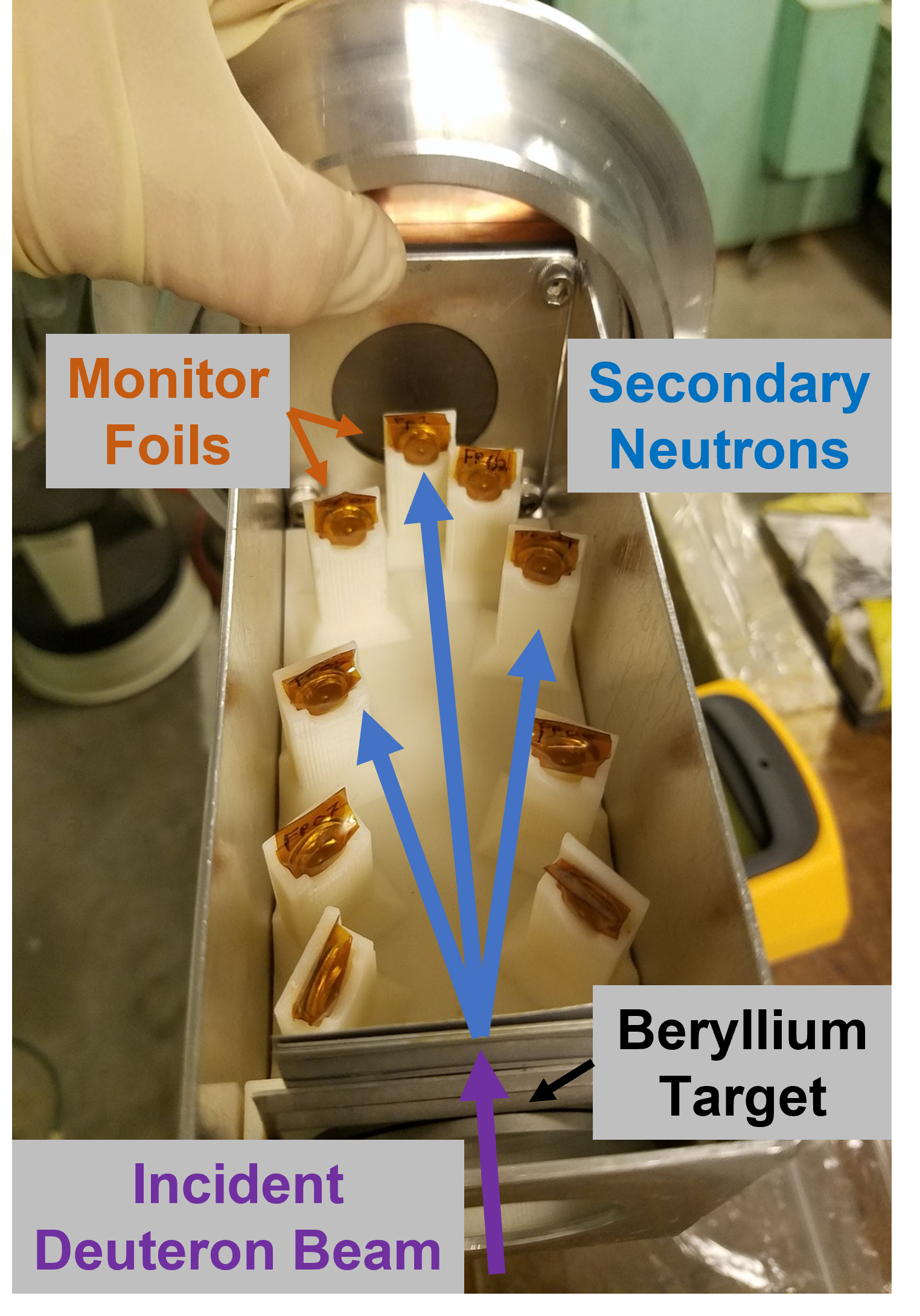}
\caption{Photo of the foil holder used in the activation experiments.  The plastic holder was loaded directly inside of the beam-pipe.  Foils were wrapped in Kapton polyimide tape for encapsulation during handling.  Thin aluminum foils were placed just behind the breakup target to prevent secondary protons from activating the samples.
}
\label{fig:foil_holder}
\end{figure}

A custom foil holder was designed and 3D-printed for the purpose of these measurements, with pre-arranged slots for the monitor foil packets at 9 different angles (see Fig. \ref{fig:foil_holder}). The beryllium breakup targets were also loaded inside of the holder, with thin aluminum sheets behind the beryllium to prevent activation from any secondary protons produced in (d,p) breakup reactions.  This entire assembly was placed into an aluminum target box, which is held under vacuum at the end of the Cave 0 beamline.  At 40 MeV, a slightly larger foil holder was printed to accommodate the additional 2 mm of beryllium required to stop the deuteron beam, which marginally changed the average angles of each foil packet.

The 33 MeV irradiation was performed with a beam current of 125 nA for 1 hour and 20 minutes, whereas the 40 MeV irradiation was performed at 75 nA for 1 hour.  Following irradiation, the foil packets were quickly removed from the beamline and transferred to the previously described HPGe counting lab. Due to the large number of foils to be counted on a single detector, and the short half-lives involved with some of the monitor reaction channels, it was decided to count the foils together in packets, rather than individually, although individual counts were performed later in time.  Counting the foils in packets meant that there was a need to correct for the photon attenuation caused by the additional foils between a particular sample and the detector.  If a given activity $A_{meas}$ was measured at time $t$, the attenuation corrected activity $A(t)$ is determined by the equation:

\begin{equation}
A(t) = A_{meas}(t) \Big( \frac{1-e^{-\mu_0\cdot x_0}}{\mu_0 \cdot x_0} \cdot \sum_{i=1}^N e^{-\mu_i \cdot x_i} \Big)^{-1} ,
\end{equation}

\noindent where the various $\mu_i$'s are photon attenuation coefficients  \cite{hubbell1995tables}, the $1-\text{exp}(-\mu_0 \cdot x_0)$ factor accounts for the attenuation within the radiogenic sample itself, and the other $N$ $\text{exp}(-\mu_i \cdot x_i)$ terms correct for the attenuation in other samples between a given sample and the detector.

The Bateman equations \cite{Bateman}, as implemented in the Curie code \cite{curie}, were used to determine the measured production rates $R_{meas,i}$ in each monitor reaction channel from these (corrected) activities, which were measured at multiple time points after the end-of-bombardment (EoB).  These measured production rates were then used to iteratively solve for the flux distribution from breakup, as previously described.  Peak-fitting of the measured HPGe spectra was also performed using the Curie code \cite{curie}.

\subsection{Time of Flight}

\begin{figure}[htb]
\centering
\includegraphics[width=0.9\linewidth]{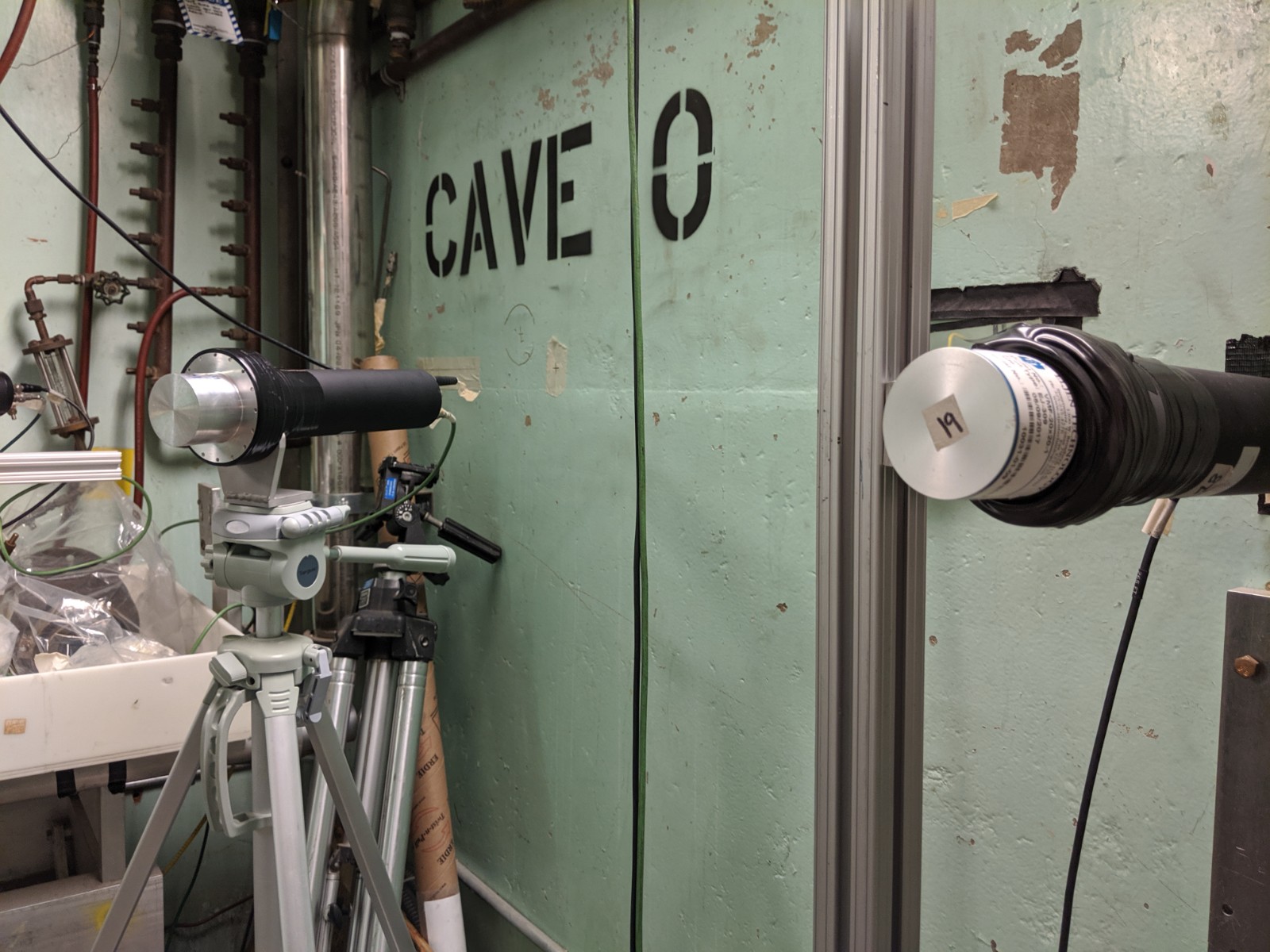}
\caption{Photo of the EJ-309 neutron scintillation detectors used in the time-of-flight experiment.  Pictured here are the detectors positioned at $0^{\circ}$ and $18.2^{\circ}$.
}
\label{fig:tof_photo}
\end{figure}

For the measurement at $\epsilon_d=40$ MeV, neutron time-of-flight (nToF) spectroscopy was performed in addition to the activation experiment, to directly measure the neutron spectrum at five individual angles.  In this case the nToF data were normalized using the 40 MeV activation measurement.  This measurement consisted of five EJ-309 liquid scintillation detectors placed at angles ranging from 0--90$^{\circ}$, which were located 1.2--2.3 m from the deuteron breakup target, with the 90$^{\circ}$ detector being placed the closest.  A photo of the 0$^{\circ}$ and 18.2$^{\circ}$ detectors can be seen in Fig. \ref{fig:tof_photo}.  The digital data acquisition system used to measure the neutron spectra from the detectors was a Mesytec MDPP-16, running the QDC (charge sensitive) firmware and the MVME recording software \cite{mesytec}.  The cyclotron RF clock frequency was also routed into one of the MDPP-16 inputs, to allow for timing the scintillation event against the cyclotron, with a window width of about 98 ns.  With this firmware, the MDPP-16 integrates the entire charge in each detected pulse, as well as the charge in a ``short" gate pulse, which together can be used to perform pulse-shape discrimination (PSD) to differentiate neutron and photon signals in the detector.

The breakup target used for the neutron time-of-flight experiment was a 6 mm thick beryllium sheet, approximately 2.25-inches square, clamped into a thick aluminum beamstop.  Because the beamstop was made from relatively thick aluminum, an attenuation correction was applied to the measured spectra at each angle using the $\ce{^{27}Al}$(n,tot) cross section from ENDF/B-VII.1 \cite{chadwick2011endf}.  The 40 MeV nToF irradiation was performed at a deuteron beam current of 0.7 nA for a duration of approximately 30 minutes, which yielded enough fluence in each detector to achieve good statistics (greater than 10,000 counts in each TDC bin).

The time-of-flight spectrum of the neutrons was determined using cuts on the integrated tail/total PSD parameter. This was done to select only events corresponding to H(n,elastic) interactions in the scintillator, as the C(n,$\alpha$) contribution to the spectrum was clearly visible, although easily separable.  There as also a low-energy cut, to eliminate the portion of the energy spectrum with poor PSD.  The time-of-flight spectrum was then converted into an energy spectrum using the relativistic time-of-flight equations:

\begin{align}
E_n(t) =& \ \ (m_n c^2)(\gamma -1) \nonumber \\
 =& \ \ (m_n c^2)\Big(\frac{1}{\sqrt{1-\big(\frac{1}{c} d/(t+\Delta t_{\gamma}) \big)^2}} -1\Big) ,
\end{align}

\noindent where $m_n c^2$ is the neutron rest mass energy, $t$ is the time of flight, and $\Delta t_{\gamma} = d/c$ is the time required for a photon to traverse the same distance as a neutron. This time-adjustment is required because the time of flight is determined relative to the characteristic ``gamma flash" that occurs when the cyclotron beam pulse first strikes the target.  This time-to-energy conversion was performed on a bin-by-bin basis, which is why the resulting spectrum (seen in Fig. \ref{fig:measured_tof}) shows non-symmetric energy bins.

In order to properly determine the neutron yields at each angle position, the detection efficiency as a function of neutron energy must be determined \cite{laplace2021}.  Because we are normalizing the nToF results with the activation measurements, we are only concerned with the relative detection efficiency.  EJ-309 detectors are composed of a scintillating fluid that is chemically similar to xylene: (CH$_3$)$_2$C$_6$H$_4$ \cite{ej309}.  At the energies measured in this experiment, neutrons primarily generate light in the detector through elastic scattering of hydrogen, or through C(n,$\alpha$) reactions.  Because the (n,$\alpha$) contribution was separated from our spectrum with PSD, we can attribute all of the neutron signal to $^1$H(n,elastic) reactions.  There is a lower limit, $E_{cut}$, below which the recoil proton will not be detected.  In this case, the value of $E_{cut}$ was set by the energy below which the PSD became poor, and was determined separately for each detector.  If we make the assumption of a small detector volume relative to the neutron path length (i.e. no multiple scattering), the detection efficiency is proportional to:

\begin{equation}
\epsilon(E_n) \propto \sigma_{el}(E_n) \cdot \frac{E_n-E_{cut}}{E_n} ,
\end{equation}

\noindent where $\sigma_{el}$ is the $^1$H(n,elastic) cross section and the factor $(E_n-E_{cut})/E_n$ arises from the fact that the neutron elastic scattering kernel for protons is uniform in energy, or more simply that the incident neutron has an equal probability of scattering to any proton energy.  
  
One issue that tends to arise with this determination of the detection efficiency is that in the low-energy portion of the spectrum, the efficiency is asymptotically dependent on the value of $E_{cut}$.  Thus the uncertainty in $E_{cut}$ gets propagated into a very large systematic uncertainty in the efficiency.  Because of this large uncertainty, the reported spectra will not include data within 0.2 MeV of $E_{cut}$, which ranged from approximately 1--3 MeV depending on the distance each detector was from the source.  A more careful consideration of the light yield from (n,p) in each detector, and a more accurate efficiency calculation would have allowed better sensitivity to low energy neutrons, and a lower $E_{cut}$.  However, because the focus of this work is on the stripping contribution to breakup at higher neutron energy, with an emphasis on applications, the 1--3 MeV cutoff was considered acceptable.

\section{Experimental Results}

The measured deuteron breakup spectra showed relatively good agreement with the predictions of the hybrid Serber model, which were based on a fit to literature data on beryllium.  The discrepancy between the Meulders 33 MeV measurement and the Saltmarsh 40 MeV measurement was confirmed, with the values observed in this experiment being very close to the measurements by Saltmarsh.  These also represent the first set of measurements of 33 MeV deuteron breakup taken at multiple angles, which will be a valuable tool for the optimization of isotope production pathways which make use of thick target deuteron breakup as a neutron source.

\begin{figure*}
    \sloppy
    \centering
    \subfloat{
        \centering
        \subfigimg[width=0.496\textwidth]{}{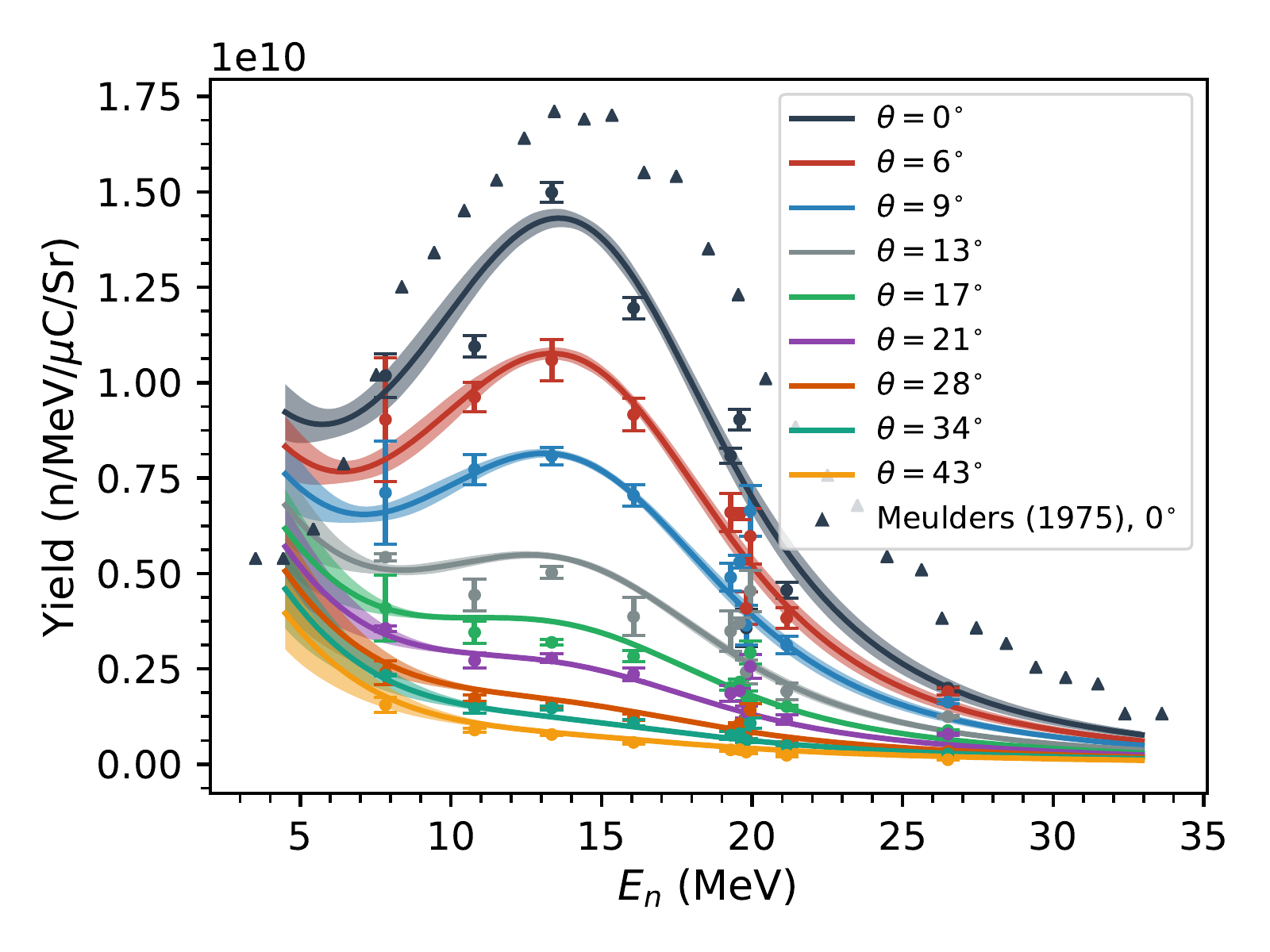}{50}
        \subfigimg[width=0.496\textwidth]{}{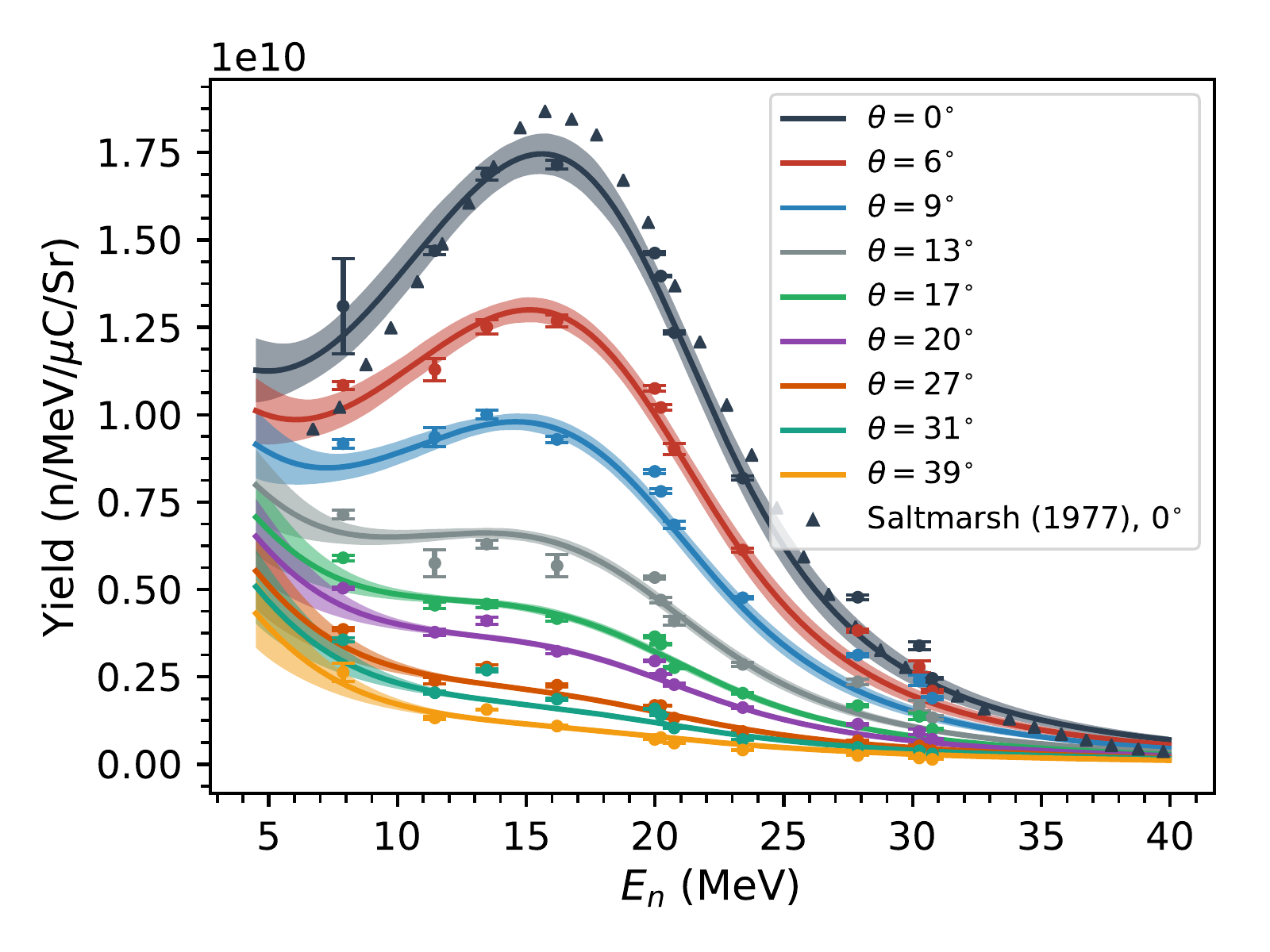}{50}
   \hspace{-10pt}}
    \caption{Results of the activation measurements at $\epsilon_d=33$ MeV (left) and $\epsilon_d=40$ MeV (right).  The centroids of the experimental data points are derived from the energy average of the cross section, i.e., $\frac{\langle E \cdot \sigma(E)\rangle}{\langle \sigma(E) \rangle}$.  The solid line (and the 1-$\sigma$ error band) indicates the hybrid Serber model fit to the data.
    }
   	\label{fig:activation_results}
\end{figure*}

\subsection{Activation Spectral Reconstruction}

The results of the foil activation experiments can be seen in Fig. \ref{fig:activation_results}.  The data show a clear agreement with the Saltmarsh measurements, and the predictions of the hybrid Serber model, and also show that the Meulders 33 MeV measurement was systematically high.  However, the measured values are approximately 12\% higher than the predictions of the hybrid model. At forward angles for 33 MeV deuterons, this is likely attributable to an underprediction of the total deuteron breakup cross section, albeit only slightly.  These measurements were determined by fitting the predicted production rates, based on the model parameters $c_1, \ldots , c_6$ in Eq. (\ref{eq:fitting}), to the measured production rates in each monitor reaction channel.  The optimized values of $c_i$ for each measurement are summarized in Table \ref{table:activation_results}.

\begin{table}[h]
\centering
\begin{tabular}{ccc}
\hline \hline
$i$ & $c_i$(33 MeV) & $c_i$(40 MeV) \Tstrut \Bstrut \\
\hline
\Tstrut  1 & 1.46 & 1.56 \\ 
2 & 0.43 & 1.93 \\ 
3 & 1.8 & 1.73 \\ 
4 & 1.1 & 1.16 \\
5 & 1.19 & 0.89 \\
6 & 0.59 & \hspace{0.08em} 0.51 \Bstrut \\
\hline \hline
\end{tabular} 
\caption{Fitted parameters for the double-differential neutron production cross section, as described in Eq. (\ref{eq:fitting}), for the 33 and 40 MeV activation experiments.}
\label{table:activation_results}
\end{table}

There are a few clear similarities between the results of the two measurements.  The angle and particularly the energy distributions for each were wider than the optimized hybrid model.  There was also a clear decrease in the pre-equilibrium contribution to the spectrum.  This could be a real, significant decrease; however, because pre-equilibrium is already a relatively small contributor to the spectrum, it could also be a statistical anomaly due to the fitting procedure.  And while the magnitude component of the total breakup cross section increased in both cases, the slope factor also changed by a factor of two in both cases (in opposite directions), so it is difficult to say whether or not this is indicative of a significant discrepancy to the ``recommended" model parameters, or the result of overfitting resulting in a local solution.  Most of these fitted parameters are likely not valid outside the scope of the data they were fitted to (i.e. should not be used for extrapolation). However, because the width of the measured energy distribution (parameter $c_3$) is consistently wider than the hybrid breakup model prediction, we can conclude that either the Meulders 50 MeV measured distributions were artificially narrower than reality, or that the hybrid model poorly predicts the deuteron energy dependence of the widths of the neutron energy distributions.

\subsection{Time of Flight Results}

\begin{figure*}
    \sloppy
    \centering
    \subfloat{
        \centering
        \subfigimg[width=0.496\textwidth]{}{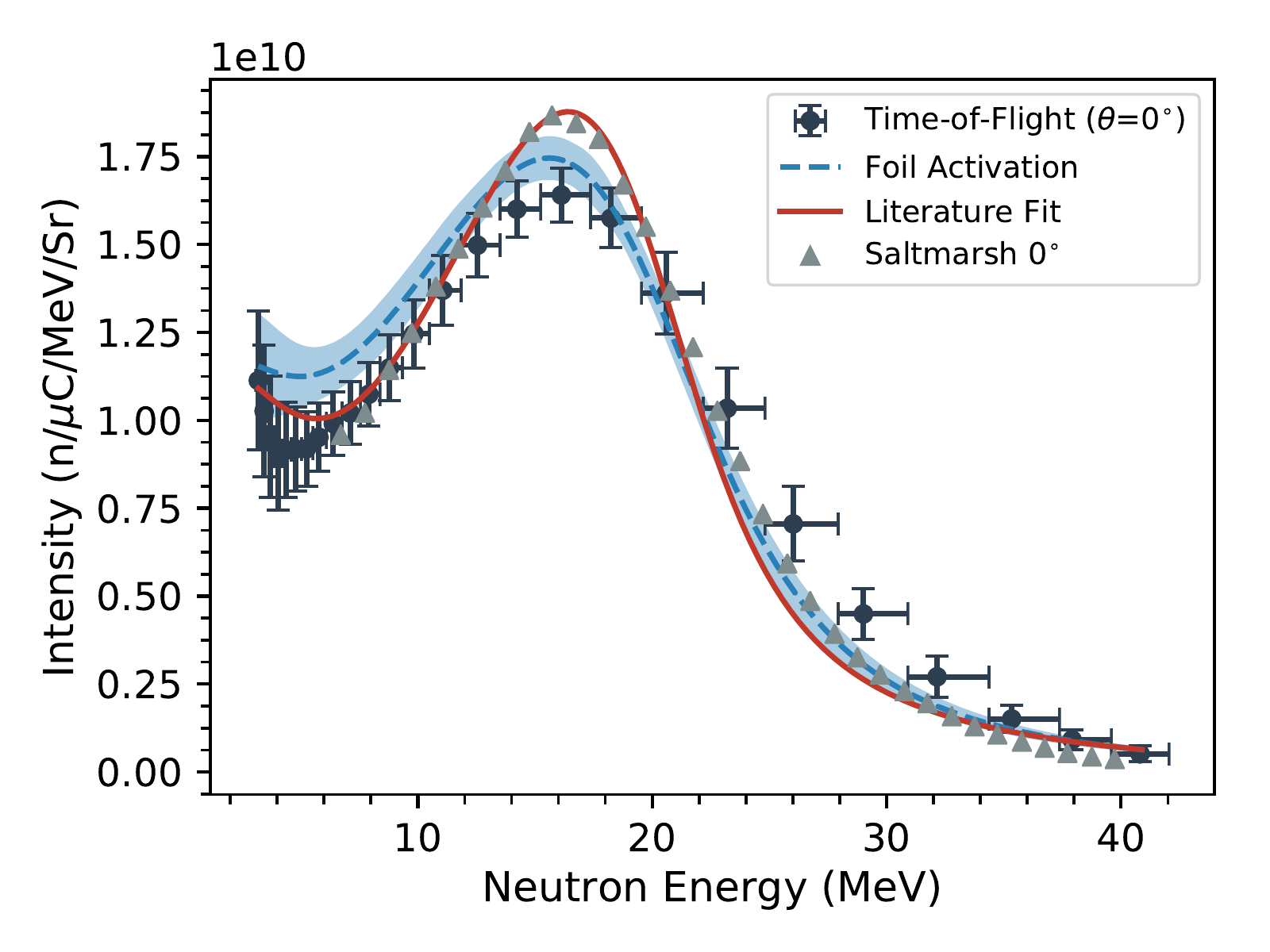}{50}
        \subfigimg[width=0.496\textwidth]{}{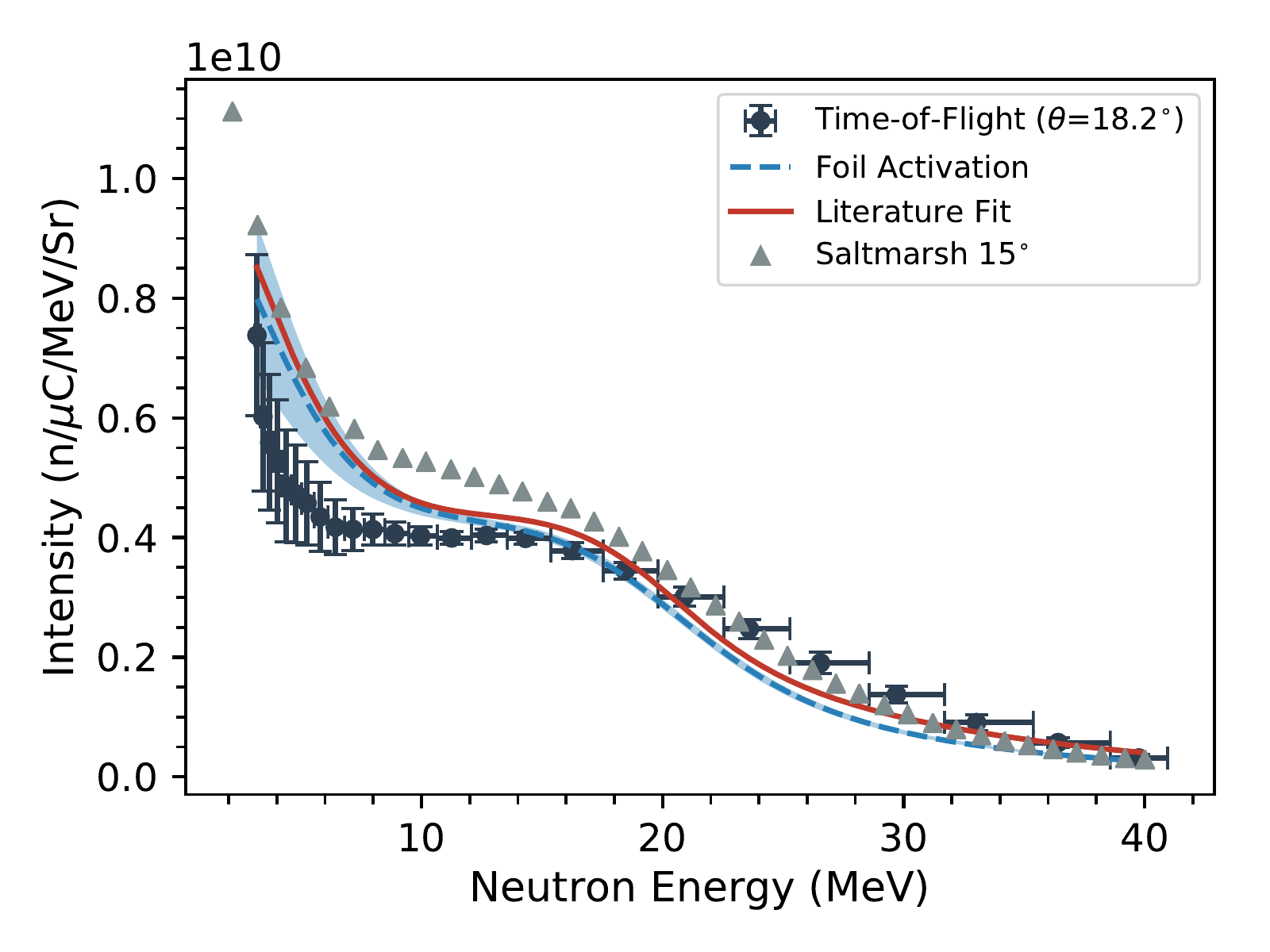}{50}
   \hspace{-10pt}}
    \\
    \subfloat{
        \centering
        \subfigimg[width=0.496\textwidth]{}{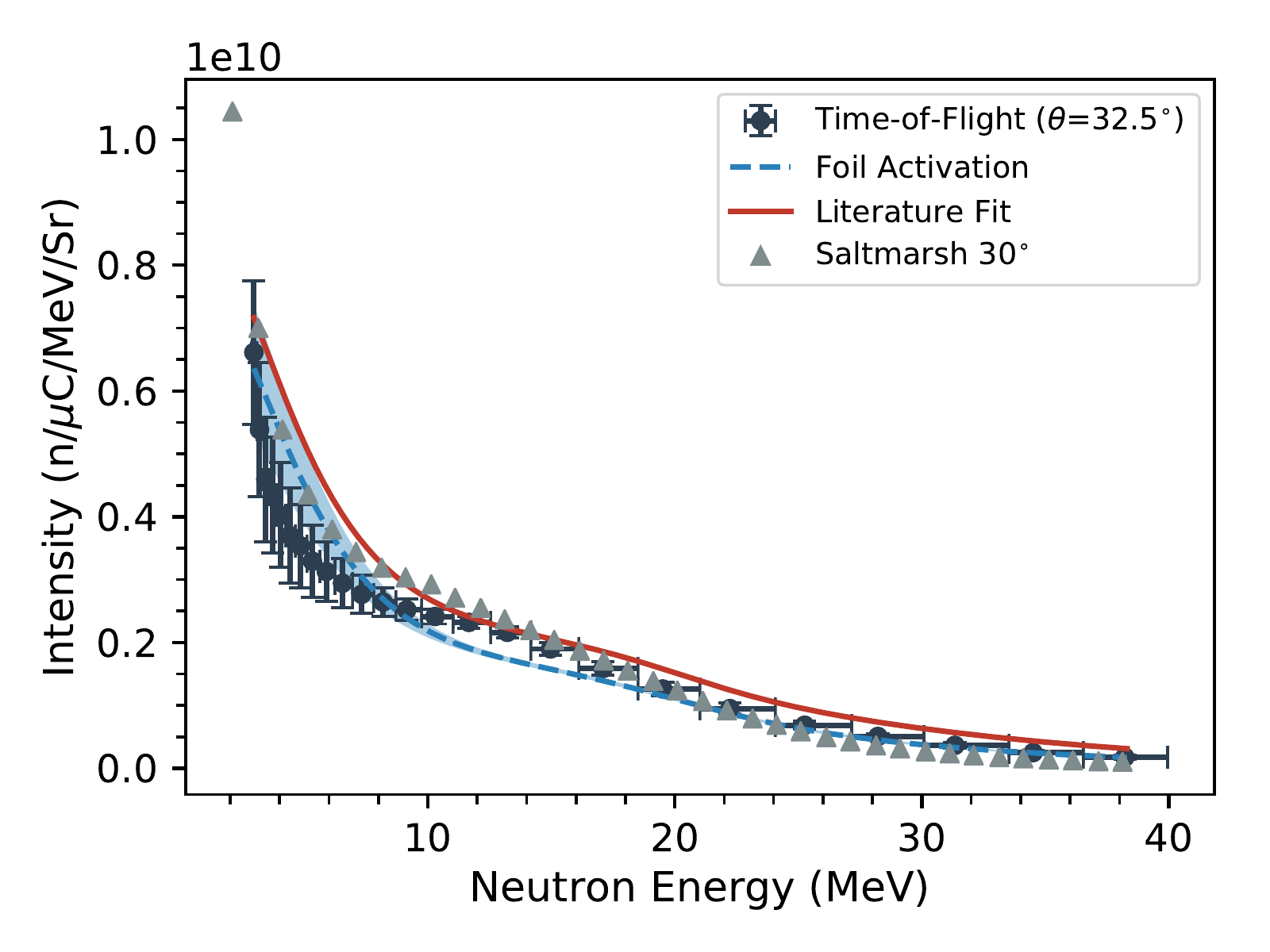}{50}
        \subfigimg[width=0.496\textwidth]{}{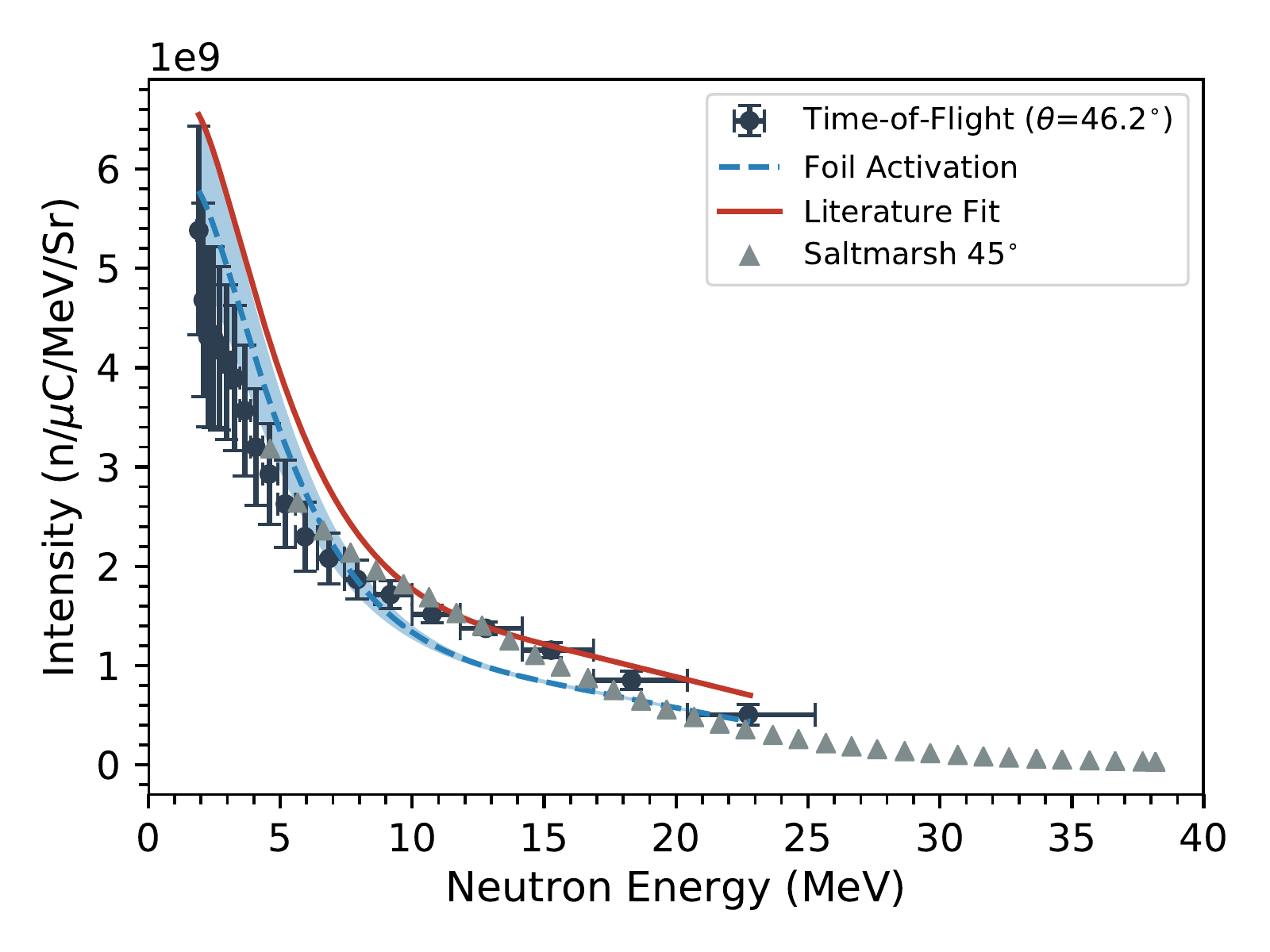}{50}
   \hspace{-10pt}}%
    \\
    \subfloat{
        \centering
        \subfigimg[width=0.496\textwidth]{}{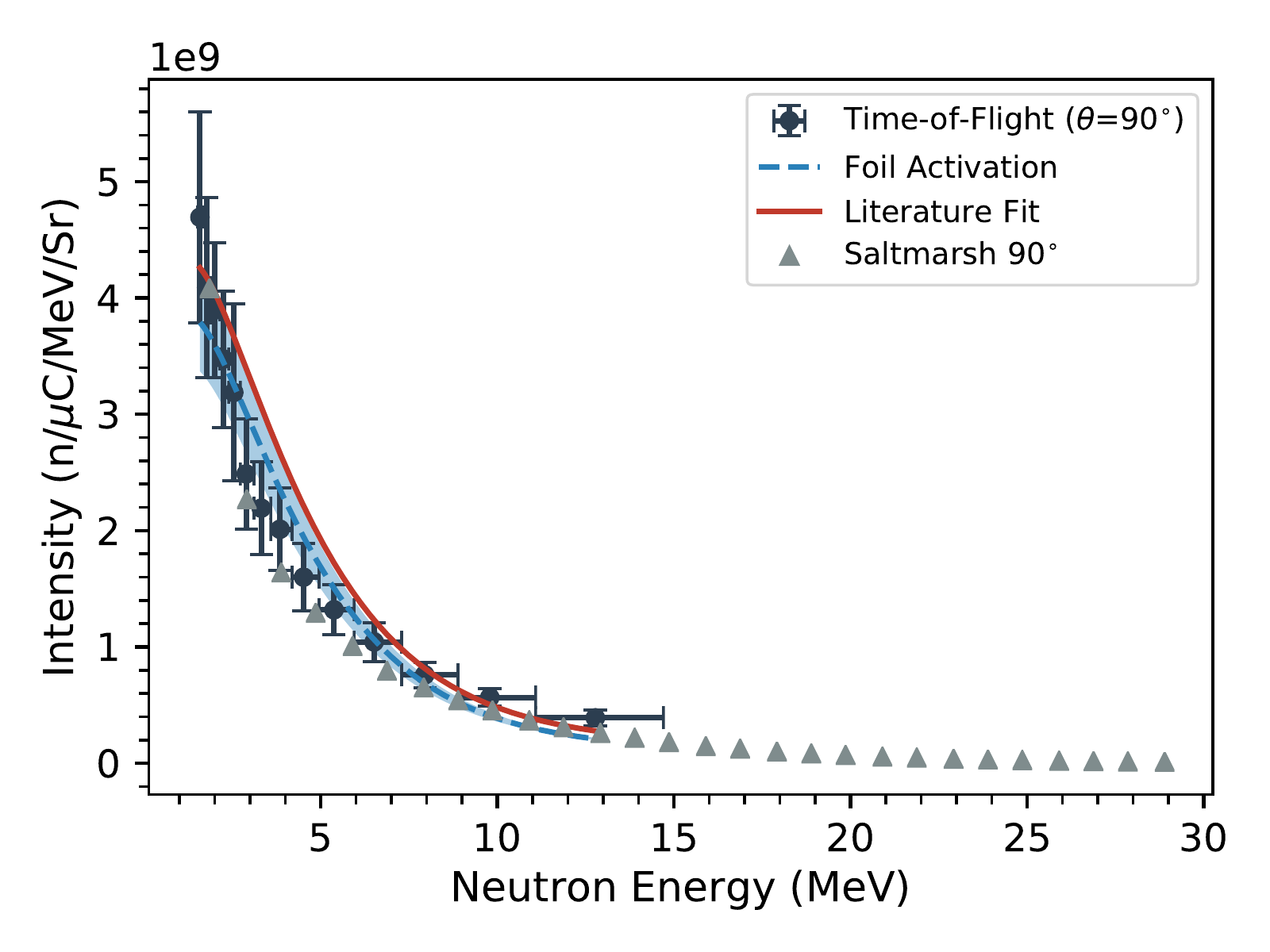}{50}
        \subfigimg[width=0.496\textwidth]{}{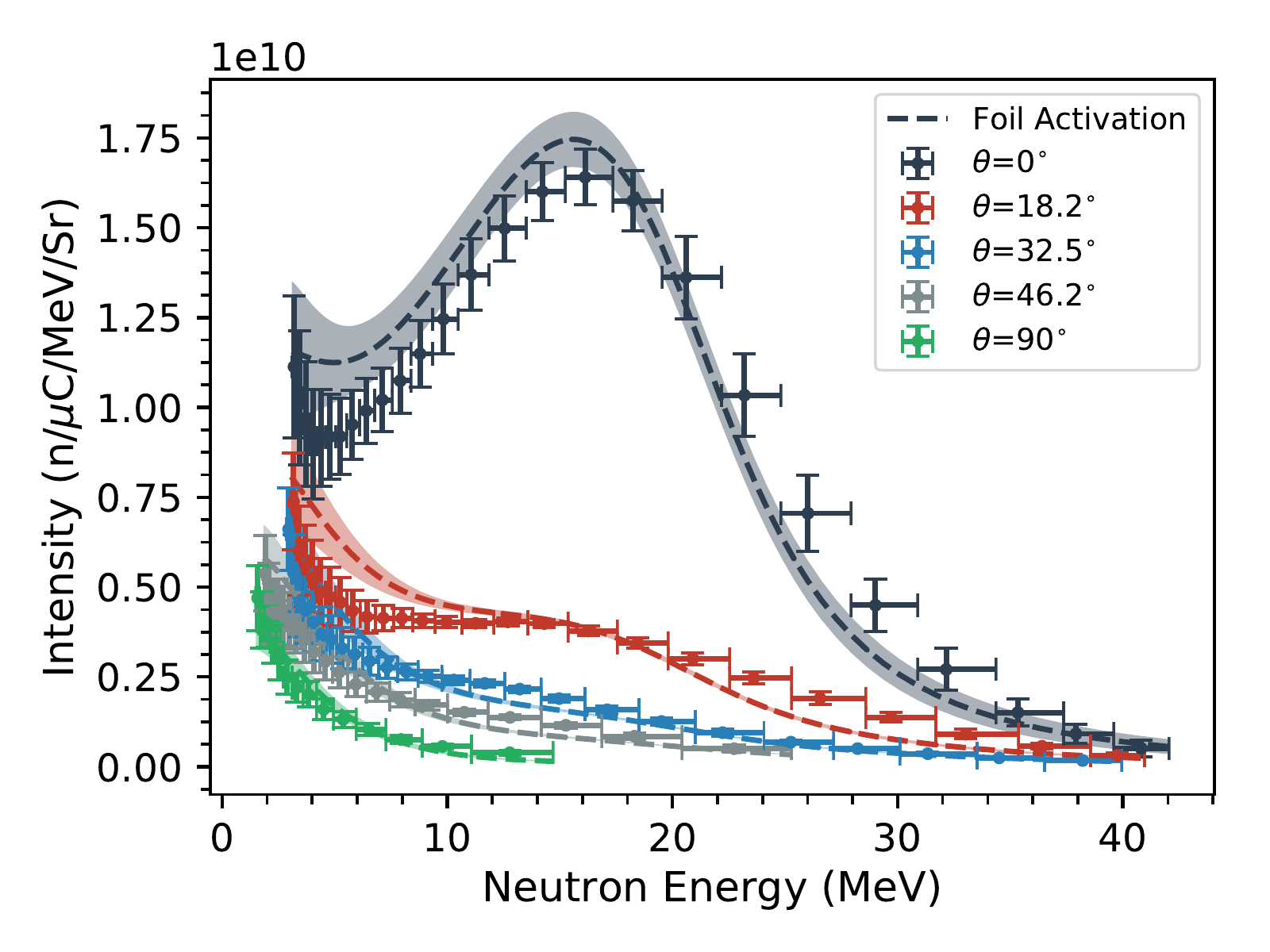}{50}
   \hspace{-10pt}}
    \caption{Measured time-of-flight spectra at $\epsilon_d = 40$ MeV.  Measured angles were $0^{\circ}$, $18.2^{\circ}$, $32.5^{\circ}$, $46.2^{\circ}$ and $90^{\circ}$.  Each plot compares the measured results to the extracted yields from the 40 MeV foil activation experiment (to which these data are normalized), our hybrid breakup model, and the 40 MeV Saltmarsh data (which were measured at the same angles, within $\approx 3^{\circ}$).  A comparison between the ToF spectra and activation data measured at all angles can be seen in the lower right.  Note that the y error-bars represent the sum of statistical error ($<1$\%), as well as the systematic error resulting from the relative efficiency determination and the normalization from the activation data.  The x ``error-bar" is being used to represent the width of each energy bin, in this case.}
    \label{fig:measured_tof}
\end{figure*}

The measured nToF results are plotted in Fig. \ref{fig:measured_tof}, along with a comparison to the 40 MeV activation results, the Saltmarsh data, and the optimized hybrid Serber model.  While the magnitudes of the nToF and activation results necessarily agree, as one was used to normalize the other, the relatively good agreement in the shape of these results does build confidence in the activation results taken at 33 MeV.  Interestingly, the increased width of the breakup neutron energy distribution, which was seen in the activation data, is also present in the nToF data.  The breakup peak is fully apparent only in the 0$^{\circ}$ spectrum, however even in the small peak at 18.2$^{\circ}$ it is clear that the measured nToF spectrum is wider than either the Saltmarsh values or the hybrid Serber model (which was itself fit to the Saltmarsh data).  

One clear area of systematic disagreement between the nToF data and the foil activation data is in the low energy portion of the spectrum, where the activation measurements are generally higher.  This could be an artifact related to wrap-around neutrons, which would overlap the high-energy side of the spectrum, leading to an under-prediction of the low-energy side of the spectrum due to the normalization procedure.  It could also be an error in the nToF efficiency determination, as the low energy portion of the spectrum, close to $E_{cut}$, will be most sensitive to that.  Or, it could be due to a lack of monitor reaction sensitivity in the low energy portion of the spectrum.  Only \ce{^{115}In}(n,n')\ce{^{115m}In} extends down to the lowest part of the spectrum, but has significant cross section up to 10 MeV.  This could be improved with more (n,n') or (n,p) reactions in this energy region, with multiple different thresholds helping to improve the sensitivity.

\subsection{Cross Section Measurements on \ce{^{nat}Zn}, \ce{^{nat}Ti} Targets}

In addition to the monitor foils, natural foils of titanium and zinc were included in the foil packs irradiated during the activation experiments.  The purpose of this was to measure the production of a number of medically significant radionuclides.  

Theranostics (therapeutic-diagnostics) are a relatively new class of radiopharmaceutical, in which both therapeutic and diagnostic isotopes (usually) of the same element are given as part of a treatment \cite{voyles2017measurement}.  Because the two isotopes are indistinguishable from a chemical perspective, they will exhibit nearly identical biochemistry.  If one of the isotopes is a positron-emitter, enabling the use of the high-resolution positron-emission tomography (PET) scanning technique, then an exact dose profile of the therapeutic isotope can be mapped through modeling of its decay processes with a similarly high resolution \cite{bailey2021developing}.  

The theranostic pairs of interest for production from zinc and titanium targets are \ce{^{64}Cu}/\ce{^{67}Cu} and \ce{^{44}Sc}/\ce{^{47}Sc}, respectively.  While both of the PET emitting isotopes \ce{^{64}Cu} and \ce{^{44}Sc} can be produced in relatively large abundances through \ce{^{63}Cu}(n,$\gamma$) and \ce{^{44}Ca}(p,n) (on an enriched target), the therapeutic $\beta^-$ emitters \ce{^{47}Sc} and \ce{^{67}Cu} lack well-established production pathways.

In the activation experiments, the production of all four of these theranostic isotopes was observed, as well as a number of neighboring reaction channels.  Unfortunately, the induced activities were measured with relatively poor statistics, mostly due to an overwhelming $\gamma$-ray background from some of the other, much stronger monitor reaction channels.  However, the measured production rates were converted into cross sections using Eq. (\ref{eq:activation}), with the model parameters for $\phi(E_n)$ from the activation experiment fits reported in Table \ref{table:activation_results}.

\begin{table}
\centering
\begin{tabular}{cc}
\hline \hline
Reaction & Multiplier \Tstrut \Bstrut \\
\hline
\Tstrut \ce{^{nat}Zn}(n,x)\ce{^{67}Cu} & 0.81 $\pm$ 0.08 \\
\ce{^{nat}Zn}(n,x)\ce{^{64}Cu} & 0.96 $\pm$ 0.09 \\
\ce{^{nat}Ti}(n,x)\ce{^{47}Sc} & 0.65 $\pm$ 0.06 \\
\ce{^{nat}Ti}(n,x)\ce{^{44}Sc} & 0.083 $\pm$ 0.025 \\
\ce{^{nat}Ti}(n,x)\ce{^{46}Sc} & 0.84 $\pm$ 0.08 \\
\ce{^{nat}Ti}(n,x)\ce{^{48}Sc} & 0.7 $\pm$ 0.05 \\
\ce{^{nat}Zn}(n,x)\ce{^{65}Zn} & 0.43 $\pm$ 0.04 \\
\ce{^{nat}Zn}(n,x)\ce{^{63}Zn} & \hspace{0.08em} 0.33 $\pm$ 0.06 \Bstrut \\
\hline \hline
\end{tabular} 
\caption{Results of multiplicative ``fit" to TENDL-2015 cross sections, based on the associated production rates measured in the Zn and Ti monitor foils \cite{rochman2017tendl}.}
\label{table:zn_ti_results}
\end{table}

We will present the results of a very basic ``fit" to these measurements, since these cross sections are averaged over a wide spectrum, have rather large uncertainties, and are somewhat difficult to represent.  The ``fit" is simply a constant multiplicative factor by which to scale the TENDL-2015 cross sections for each channel, such that they best reproduce the measured flux-averaged cross sections.  The optimized scaling parameters from this fit are given in Table \ref{table:zn_ti_results}.

\section{Summary and Conclusions}

In this work we have proposed a new hybrid model for predicting neutron yields from intermediate-energy deuteron breakup on light targets. The model was benchmarked against literature values at a variety of incident deuteron energies and outgoing neutron energies and angles.  This model has the benefits of a relatively simple calculation method, that does not sacrifice accuracy over the application range considered.  It has also been adapted to apply a relatively simple 6-parameter fitting procedure, which was demonstrated to be applicable to both global fits across the literature values mentioned above, as well as a more focused fit to neutron monitor activation data for spectrum unfolding.  

We have also presented new measurements of the breakup spectrum on beryllium for deuteron energies of 33 and 40 MeV.  These were performed with neutron activation unfolding, using the hybrid breakup model to forward-fit to the spectrum, as well as complementary time-of-flight measurements, which were normalized to the activation data.  The results of this showed that the hybrid model was generally accurate.  Additionally, measurements of a number of medically relevant isotope production pathways were presented, along with a recommended cross section based on these measurements.

In the future, this work could be extended by pursuing measurements at other energies or low-$Z$ materials (such as Li), to explore the extrapolation capabilities of the model, and should also incorporate direct reaction theory.  It is the hope that this work, in providing a relatively simple yet accurate model of deutron breakup on light targets, will aid in the design of future high intensity neutron sources based on this mechanism, and will enable new research into applications of intense neutron sources, such as isotope production.

\section*{Acknowledgements}
We wish to acknowledge the scientists, engineers, operators, and support staff of the 88-Inch Cyclotron, who enabled the successful completion of the experimental portions of this work.  In particular, we would like to thank the operators Brien Ninemire, Nick Brickner, Scott Small, and Devin Thatcher for providing a well-tuned deuteron beam.  We would also like to thank Juan Manfredi and Thibault Laplace for their expertise in setting up the neutron detectors and data acquisition hardware for the neutron time-of-flight portion of this work.

This research is supported by the U.S. Department of Energy Isotope Program, managed by the Office of Science for Isotope R\&D and Production, carried out under Lawrence Berkeley National Laboratory (Contract No. DE-AC02-05CH11231), and with additional support from the U.S. Nuclear Data Program.

\bibliography{references}

\newpage

\ \

\appendix

\subsection{Relevant Nuclear Data}
\label{nudat_appendix_deuteron}

\begin{table}[H]
\centering
\begin{tabular}{cccc}
\hline\hline
Isotope & $\gamma$ Energy (keV) & $I_{\gamma}$ (\%) & $T_{1/2}$ \Bstrut \\
\hline
\Tstrut
\ce{^{24}Na} & 1368.626 & 99.9936 (15) & 3.16 (4) d \\
 & 2754.007 & 99.855 (5) & \\
\ce{^{44}Sc} & 1157.02 & 99.9 (4) & 3.97 (4) h \\
\ce{^{46}Sc} & 889.277 & 99.984 (1) & 83.79 (4) d \\
 & 1120.545 & 99.987 (1) & \\
\ce{^{47}Sc} & 159.381 & 68.3 (4) & 3.3492 (6) d \\
\ce{^{48}Sc} & 175.361 & 7.48 (1) & 43.67 (9) h \\
 & 983.526 & 100.1 (6) & \\
 & 1037.522 & 97.6 (7) & \\
 & 1212.88 & 2.38 (4) & \\
 & 1312.12 & 100.1 (7) & \\
\ce{^{57}Co} & 122.06065 & 85.6 (17) & 271.74 (6) d \\
 & 136.47356 & 10.68 (8) & \\
\ce{^{57}Ni} & 127.164 & 16.7 (5) & 36.60 (6) h \\
 & 1377.63 & 81.7 (4) & \\
 & 1919.52 & 12.3 (4) & \\
\ce{^{58}Co} & 810.7593 & 99.45 (1) & 70.86 (6) d \\
\ce{^{63}Zn} & 669.62 & 8.2 (3) & 38.47 (5) m \\
 & 962.06 & 6.5 (4) & \\
\ce{^{64}Cu} & 1345.77 & 0.475 (11) & 12.701 (2) h \\
\ce{^{65}Zn} & 1115.539 & 50.04 (1) & 243.93 (9) d \\
\ce{^{67}Cu} & 93.311 & 16.1 (2) & 61.83 (12) h \\
 & 184.577 & 48.7 (3) & \\
\ce{^{87m}Y} & 380.79 & 78.05 (8) & 13.37 (3) h \\
\ce{^{87}Y} & 388.5276 & 82.2 (11) & 79.8 (3) h \\
 & 484.805 & 89.8(9) & \\
\ce{^{88}Y} & 898.042 & 93.7 (3) & 106.626 (21) d \\
 & 1836.063 & 99.2 (3) & \\
\ce{^{89}Zr} & 909.15 & 99.04 (5) & 78.41 (12) h \\
\ce{^{111}In} & 171.28 & 90.7 (9) & 2.8047 (4) d \\
 & 245.35 & 94.1 (1) & \\
\ce{^{113m}In} & 391.698 & 64.94 (17) & 99.476 (23) m \\
\ce{^{114m}In} & 190.27 & 15.56 (15) & 49.51 (1) d \\
 & 558.43 & 4.4 (6) & \\
 & 725.24 & 4.4 (6) & \\
\ce{^{115m}In} & 336.241 & 45.9 (1) & \hspace{0.08em} 4.486 (4) h \\
\hline\hline
\end{tabular}
\label{table:decay_data}
\caption{Principle $\gamma$-ray data from ENSDF \cite{A24, A44, A46, A47, A48, A57, A58, A63, A64, A65, A67, A87, A88, A89, A111, A113, A114, A115}.  Uncertainties are listed in the least significant digit, that is, 3.16(4) d means 3.16 $\pm$ 0.04 d.}
\end{table}
\ \

\end{document}